\documentclass[twocolumn,prb]{revtex4-1}
\usepackage{epsfig,amssymb,amsmath}
\usepackage{graphicx}
\usepackage{color}

\begin{document}
\title{Ferromagnetic resonance and magnetic precessions in $\varphi_0$ junction}
\author{Yu. M. Shukrinov~$^{1,2}$}
\author{I. R. Rahmonov~$^{1,3}$}
\author{K. Sengupta~$^{4}$}

\address{$^{1}$ BLTP, Joint Institute for Nuclear Research, Dubna, Moscow Region, 141980, Russia\\
$^{2}$ Dubna State University, Dubna,  141980, Russia\\
$^{3}$ Umarov Physical Technical Institute, TAS, Dushanbe, 734063, Tajikistan\\
$^{4}$ School of Physical Sciences, Indian Association for the
Cultivation of Science, Jadavpur, Kolkata-700032, India}

\date{\today}

\begin{abstract}
The Josephson $\varphi_0$  junctions with the current-phase relation $I = I_c \sin (\varphi-\varphi_0)$, where the phase
shift $\varphi_0$ is proportional to the magnetic moment
perpendicular to the gradient of the asymmetric spin-orbit potential, demonstrate a number of unique features important for superconducting spintronics and modern informational technologies. Here we show that a current sweep along IV-characteristic of the $\varphi_0$ junction may lead to regular magnetization dynamics with a series of specific phase trajectories. The origin of these trajectories is related to
a direct coupling between the magnetic moment and the Josephson
oscillations in these junctions, and ferromagnetic resonance when Josephson frequency coincides with the ferromagnetic one. We demonstrate that an external
electromagnetic field can control the dynamics of magnetic moment
within a current interval corresponding to a Shapiro step and
produce topological transformation of specific precession
trajectories. We demonstrate the appearance of the DC component of superconducting current and clarify its role in the transformation of IV-characteristics in the resonance region.
Good agreement between numerical and analytical results has been
found in the ferromagnetic resonance region. The presented results might be used for developing novel resonance methods of determination of the spin-orbit coupling parameter in the non-centrosymmetric  materials.
We discuss experiments which can test our results.
\end{abstract}

\maketitle

\section{Introduction.} Superconducting spintronics is one of the intensively developing
fields of condensed matter physics today. An important place in this field
is occupied by the investigations of Josephson junctions (JJs)
coupled to magnetic systems~\cite{linder15,efetov11}. The ability to
manipulate the magnetic properties by Josephson current and its
opposite, {\it i.e.} to influence the Josephson current by the magnetic
moment, has attracted much recent attention~\cite{buzdin05,bergeret05,golubov04,ghosh17}. The central role in these
phenomena is played by spin-orbit interaction. In the
superconductor /ferromagnet/ superconductor (S/F/S) Josephson
junctions, the spin-orbit interaction in a ferromagnet without
inversion symmetry provides a mechanism for a direct (linear)
coupling between the magnetic moment and the superconducting
current. Such noncentrosymmetric ferromagnetic junctions, called
hereafter $\varphi_0$ junctions, break time reversal symmetry.
Consequently, the current-phase relation (CPR) of these junctions
is given by $I = I_c \sin (\varphi-\varphi_0)$, where the phase
shift $\varphi_0$ is proportional to the magnetic moment
perpendicular to the gradient of the asymmetric spin-orbit potential~\cite{buzdin08}. This feature of the CPR allows one to manipulate
the internal magnetic moment using the Josephson current~\cite{buzdin08,konschelle09}.
The theory of the anomalous Zeeman effect and spin-galvanic effect in $\varphi_0$ junctions was discussed  in Refs.\ \cite{dolcini15,konshelle15}. Experimental realization of the $\varphi_0$  junction has been recently reported by Szombati et al.~\cite{szombati}. In Ref.\ \cite{chudn2016}, the authors argued that the $\varphi_0$  Josephson junction is ideally suited for studying of quantum tunneling of the magnetic moment. They proposed that magnetic tunneling would show up in the ac voltage across the junction and it could be controlled by the bias current applied to the junction.

Though the static properties of
the $S/F/S$ structures are well studied both theoretically and
experimentally, much less is known about the magnetic dynamics of
these systems~\cite{waintal02,braude08,linder83}. Recently, the
presence of an anomalous phase shift of $\varphi_0$  was
experimentally observed directly
through CPR measurement in a hybrid SNS JJ
fabricated using ${\rm Bi_2Se_3}$ (which is a topological insulator
with strong spin-orbit coupling) in the presence of an in-plane
magnetic field~\cite{aprili}. This constitutes a direct experimental
measurement of the spin-orbit coupling strength and opens up new
possibilities for phase-controlled Josephson devices made from
materials with strong spin-orbit coupling.

It was demonstrated that the DC superconducting current applied to a
$S/F/S$ $\varphi_0$ junction might produce a strong orientation
effect on the ferromagnetic layered magnetic moment~\cite{apl17}.
The application of DC voltage to the $\varphi_0$ junction would
produce current oscillations and consequently magnetic precession.
As shown in Ref.\ \cite{konschelle09}, this precession may be monitored by the appearance of higher
harmonics in the CPR as well as by the presence of a DC component of
the superconducting current that is increases substantially near the ferromagnetic
resonance (FMR). The authors stressed that the magnetic dynamics of the $S/F/S$ $\varphi_0$ junction may be quite complicated and strongly anharmonic. In contrast to these results, we demonstrate here that precession of the magnetic moment  in some current intervals along IV-characteristics may be very simple and harmonic.   It is expected that external
radiation would lead to a series of novel phenomena. Out of this, the
possibility of appearance  of half-integer Shapiro steps (in
addition to the conventional integer steps) and  the generation of
an additional magnetic precession with frequency of external
radiation was already discussed in Ref.\ \cite{konschelle09}.
However, to the best of our knowledge, an important problem related
to the reciprocal influence of Josephson current and magnetization at
different bias current  along the current-voltage
(IV)-characteristics has not been investigated till now.
Furthermore, the variation of the magnetic precessions in
the $\varphi_0$ junction along the IV-characteristics has not also been
addressed.

In this paper, we present the results on the magnetic precession in the
$\varphi_0$ junctions with a current sweep along IV-characteristic. This allows us to find specific current intervals with very simple magnetization dynamics. We show that the origin of these trajectories is related to a direct coupling between the magnetic moment and the Josephson
oscillations, realized in these junctions, and manifestation of ferromagnetic resonance features when the Josephson frequency is close to the ferromagnetic one. We also demonstrate that the interaction of
the Josephson current and the magnetic moment manifests several
interesting features under external electromagnetic radiation. In
particular, the external radiation can tune the nature
of magnetic moment precession in a current interval corresponded to
the Shapiro step. We show that such external radiation can
produce a topological transformation of magnetization precession
trajectories.  We numerically demonstrate the appearance of the dc-component of superconducting current and clarify its role in the transformation of IV-characteristics in the resonance region. The effects of Gilbert damping and spin-orbit coupling on IV-characteristics, magnetization precession and ferromagnetic resonance features are clarified.

\section{ Model and Method. }
In Josephson junctions with a thin ferromagnetic layer the superconducting phase difference and magnetization of the $F$ layer are two coupled dynamical variables. The system of equations describing the  dynamics of these variables is obtained from the Landau-Lifshitz-Gilbert equation and Josephson relations for current and phase difference.
Particularly, the magnetization dynamics of our system is described by the
Landau-Lifshitz-Gilbert equation where the effective field depends on the phase difference
\begin{eqnarray}
\frac{d {\bf M}}{dt} &=& -\gamma {\bf M} \times {\bf
H_{eff}}+\frac{\alpha}{M_{0}}\bigg({\bf M}\times \frac{d {\bf
M}}{dt} \bigg), \nonumber\\
{\bf H_{eff}} &=& \frac{K}{M_{0}}\bigg[G r \sin\bigg(\varphi - r
\frac{M_{y}}{M_{0}} \bigg) {\bf\widehat{y}} +
\frac{M_{z}}{M_{0}}{\bf\widehat{z}}\bigg], \label{llg1}
\end{eqnarray}
where $\gamma$ is the gyromagnetic ratio, $\alpha$ is a
phenomenological damping constant, $M_{0}=\|{\bf M}\|$,
$\displaystyle G= E_{J}/(K \mathcal{V})$, $K$ is an anisotropic
constant, $\mathcal{V}$ is the volume of ferromagnetic $F$ layer,  $l=4
h L/\hbar \upsilon_{F}$, $L$ is the length of the $F$ layer, and $h$
denotes the exchange field in the ferromagnetic layer.

Based on the equations for JJ and magnetic system, we can rewrite total
system of equations (to be used in our numerical studies) in
normalized units
\begin{equation}
\label{syseq}
\begin{array}{llll}
\displaystyle \dot{m}_{x}=\frac{\omega_F}{1+\alpha^{2}}\{-m_{y}m_{z}+Grm_{z}\sin(\varphi -rm_{y})\\
-\alpha[m_{x}m_{z}^{2}+Grm_{x}m_{y}\sin(\varphi -rm_{y})]\},
\vspace{0.2 cm}\\
\displaystyle \dot{m}_{y}=\frac{\omega_F}{1+\alpha^{2}}\{m_{x}m_{z}\\
-\alpha[m_{y}m_{z}^{2}-Gr(m_{z}^{2}+m_{x}^{2})\sin(\varphi -rm_{y})]\},
\vspace{0.2 cm}\\
\displaystyle \dot{m}_{z}=\frac{\omega_F}{1+\alpha^{2}}\{-Grm_{x}\sin(\varphi -rm_{y})\\
-\alpha[Grm_{y}m_{z}\sin(\varphi -rm_{y})-m_{z}(m_{x}^{2}+m_{y}^{2})]\},
\vspace{0.2 cm}\\
 \frac{d V}{d t}=\frac{1}{\beta_{c}}[I-V-\sin(\varphi-rm_{y})],
 \quad  \frac{d \varphi}{d t}=V,
 \end{array}
\end{equation}
\noindent where $\beta_{c}=2e I_{c}CR^{2}/\hbar$ is the McCumber
parameter, $m_{i}= M_{i}/M_0$ for $i=x,y,z$, and
$\omega_F=\Omega_F / \omega_c$ with the ferromagnetic resonance frequency $\Omega_F=\gamma K/M_0$ and characteristic frequency
$\omega_{c} =2eRI_{c}/\hbar$.
Here we normalize time in units of $\omega^{-1}_{c}$, external
current $I$ in units of $I_{c}$, and the voltage $V$ in units of
$V_{c}=I_{c}R$. This system of equations, solved numerically using
the fourth order Runge--Kutta method, yields $m_{i}(t)$, $V(t)$ and $\varphi(t)$ as a function of the
external bias current $I$. After averaging procedure~\cite{prb07,kleiner-book} we can find IV-characteristics at fixed system's parameters and investigate the dynamics of magnetization along the IV-curve~\cite{prb-ivc}.

The system of equations (\ref{syseq}) is significantly simplified in the
absence of dissipation ($\alpha=0$). It allows us
to clearly understand the magnetization dynamics
corresponding to the different points of the IV-curve. For this
purpose we calculate the temporal dependence of $V$ and $m_{i}$ at small dissipation ($\alpha\ll1$)
at each value of bias current. In parallel, we also verify the dynamics
of the magnetic system by solving the LLG equations at some averaged
values of voltage. The results obtained in this case are in qualitative agreement with
the solutions of the system (\ref{syseq}).

\section{ Manifestation of FMR at small $G$ and $r$.}
When the current smaller than the critical one ($I<I_c$) is applied to the $\varphi_0$ junction, the rotation of the magnetic moment $M_y$  is determined by $\sin \theta=(I_s/I_c)Gr$  ($M_y=M\sin \theta$, $\theta$ is the angle between the z-axis and the $M$ direction), which signifies that superconducting current provokes the rotation of $M$ in the $yz$-plane~\cite{konschelle09}.

First, we demonstrate the interaction between superconducting current and magnetic moment and show the manifestation of the ferromagnetic resonance in the $\varphi_0$ junction. For this purpose we calculate the $I$--dependence of the maximal $m_{y}$-component at each step of bias current. To see effect clearly, we make simulations at very small values of the parameters $G$ and $r$.

\begin{figure}[h!]
\centering
\includegraphics[width=4cm]{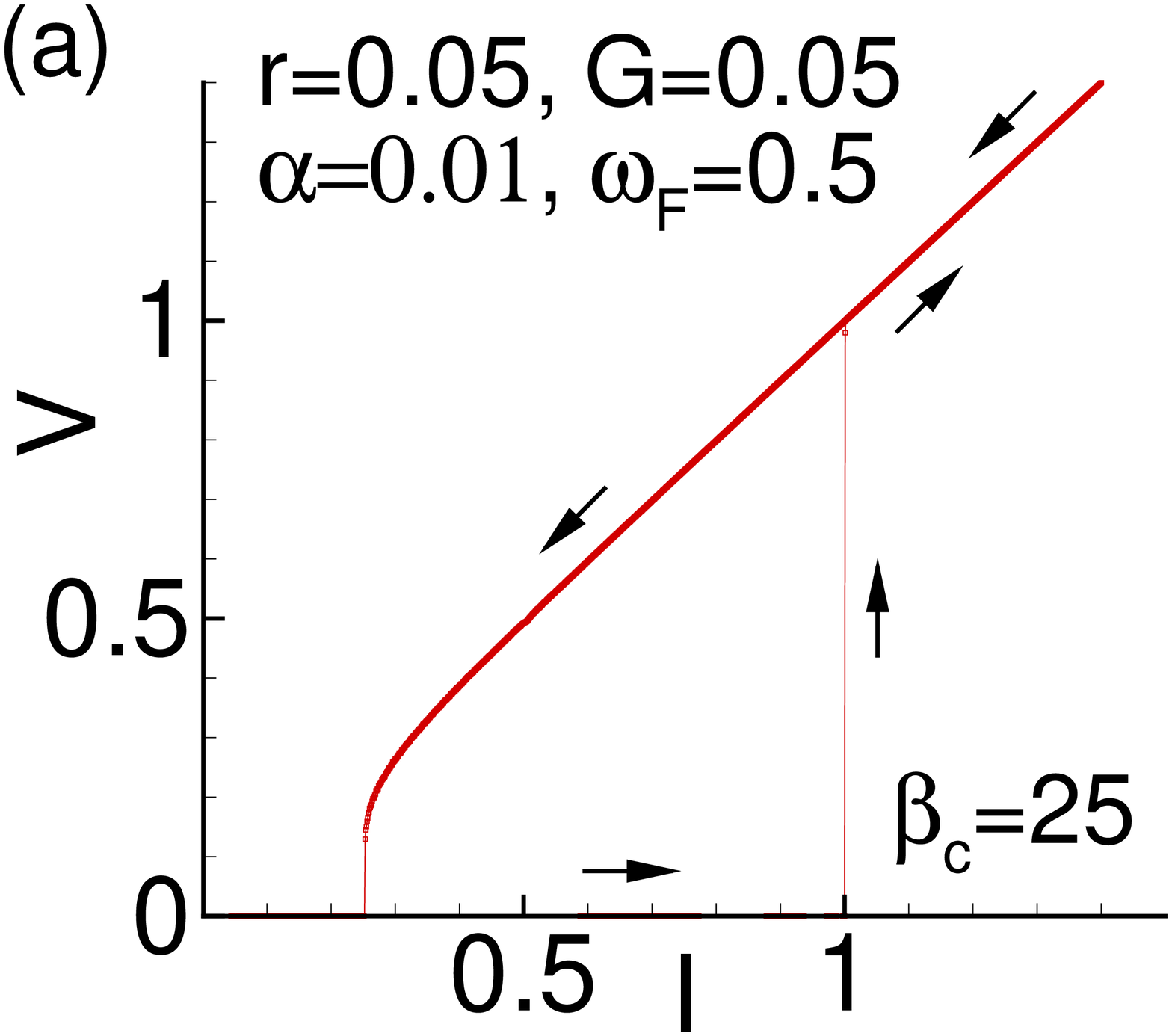}\includegraphics[width=4cm]{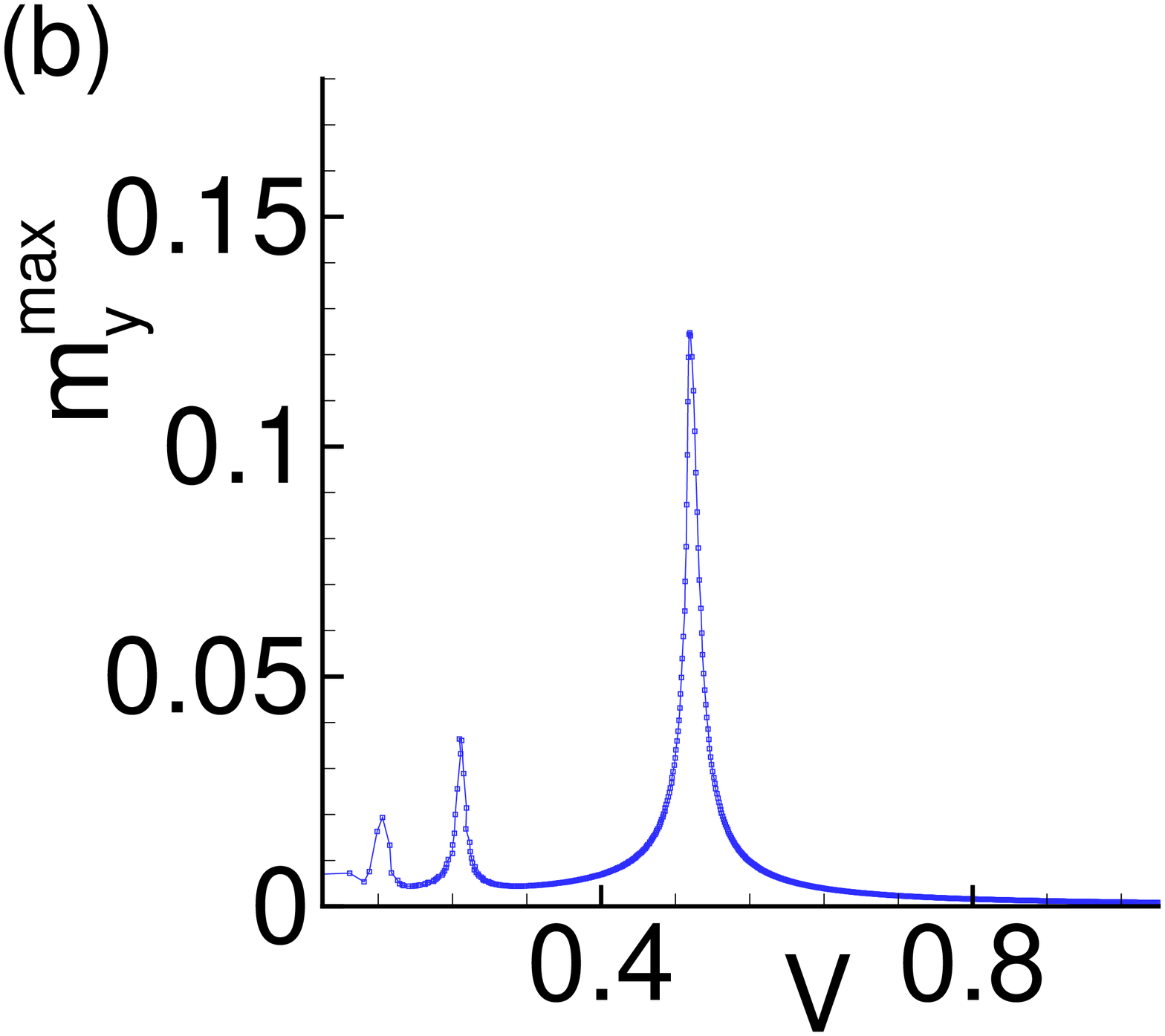}\\
\includegraphics[width=4cm]{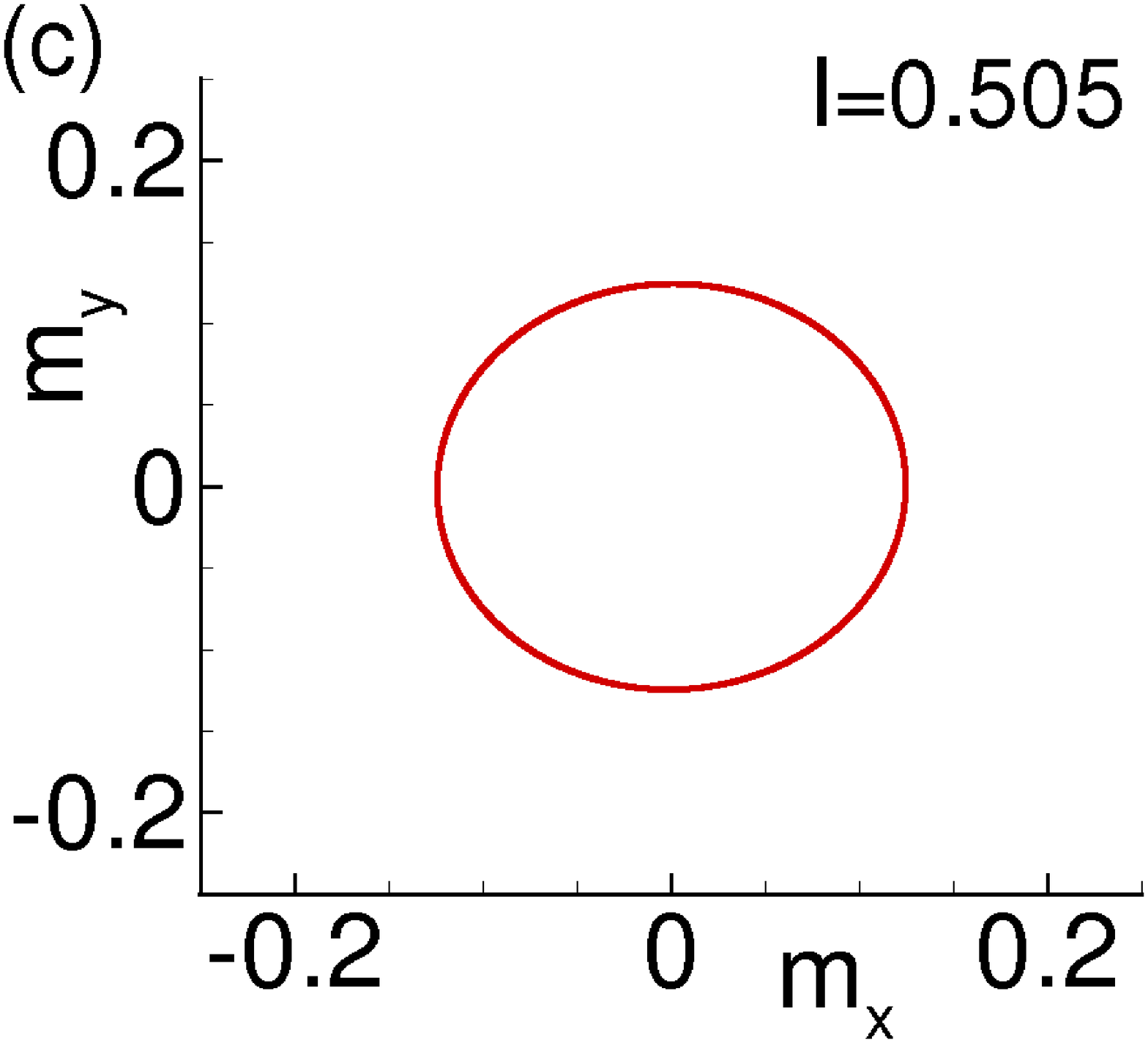}\includegraphics[width=4cm]{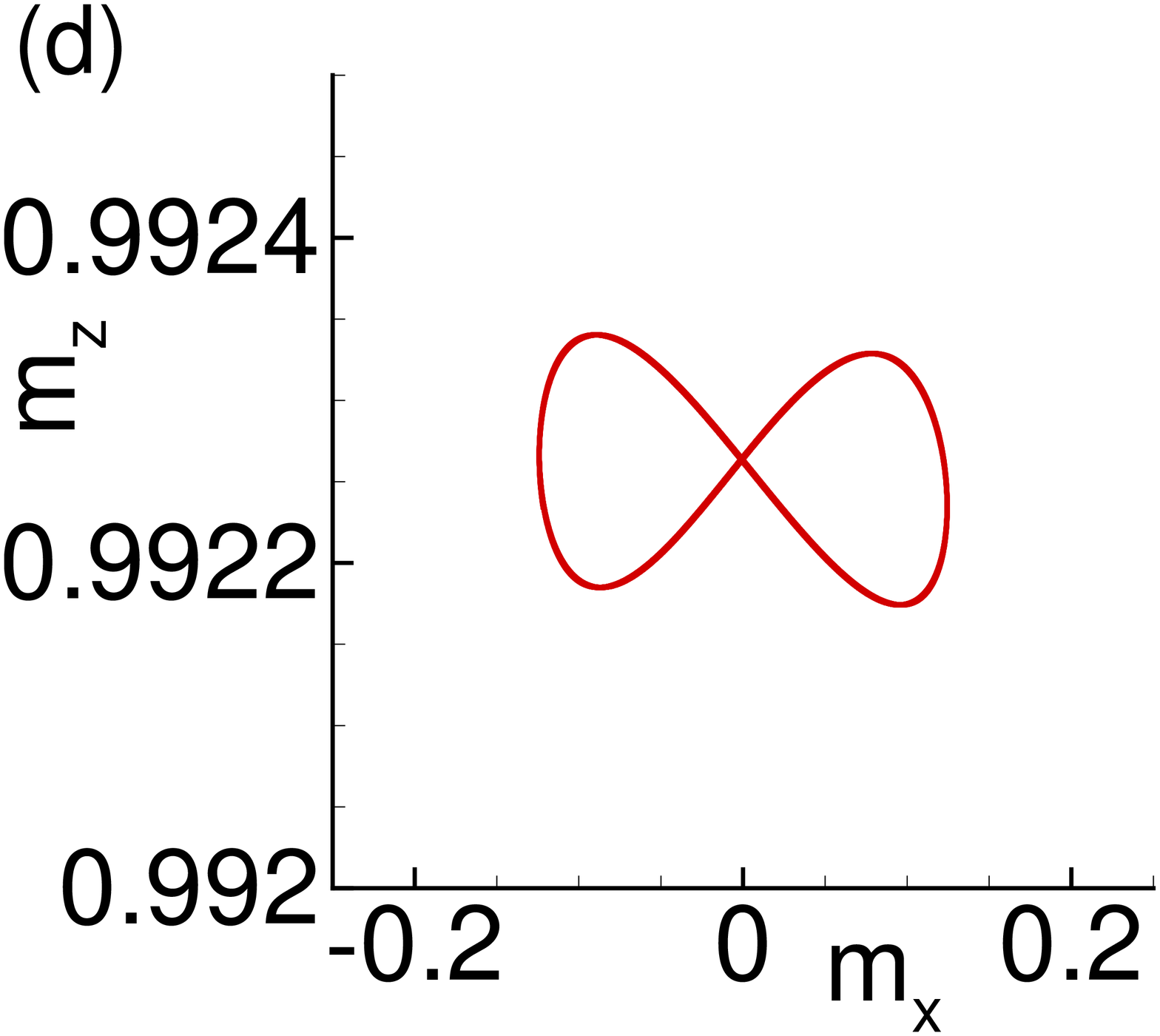}
\caption{(a) IV--characteristic of the $\varphi_{0}$--junction; (b) Manifestation of the ferromagnetic resonance in the voltage dependence of $m^{max}_{y}$; (c) Magnetization trajectory in the $m_x-m_y$ plane at current $I=0.505$ corresponded to the maximum of the resonance curve in (b); (d) The same in the $m_x-m_z$ plane.} \label{1}
\end{figure}

In Fig.\ \ref{1}(a), we present the calculated
one-loop IV-curve (obtained by increasing and decreasing $I$)  which displays an expected hysteresis for
$\beta_c=25$. The IV-characteristic at the chosen parameters of the system does not react practically on the changes in the magnetization dynamics. Later, we will discuss this question and show  the resonance manifestation in the IV-curve at larger $r$ and $G$. Here we will concentrate on the features near the ferromagnetic resonance which corresponds to the return branch around $\omega_F=V=0.5$. Figure~\ref{1}(b) presents the voltage dependence of the maximal amplitude of magnetic moment oscillations  $m^{max}_y$ taken in the time domain at each value of bias current calculated at small values of the Josephson to magnetic energy relation $G=0.05$ and the parameter of spin-orbit coupling $r=0.05$. We see the resonance peak around $V=\omega_J=\omega_F=0.5$ and its two harmonics at $V=\omega_J/2$ and  $V=\omega_J/3$.  The trajectory of magnetization in the $m_y-m_x$ plane at bias current $I=0.505$ corresponded to the resonance peak is shown in Fig.~\ref{1}(c). The dynamics of magnetization is very simple here and corresponds to the rotation of the magnetic moment around the $z$ axis. A small deviation to the $y$ axis change periodically during one rotation circle, so in the $m_z-m_x$ plane the magnetic moment describes the form shown in Fig.\ref{1}(d).

\begin{figure}[h!]
\centering
\includegraphics[width=8cm]{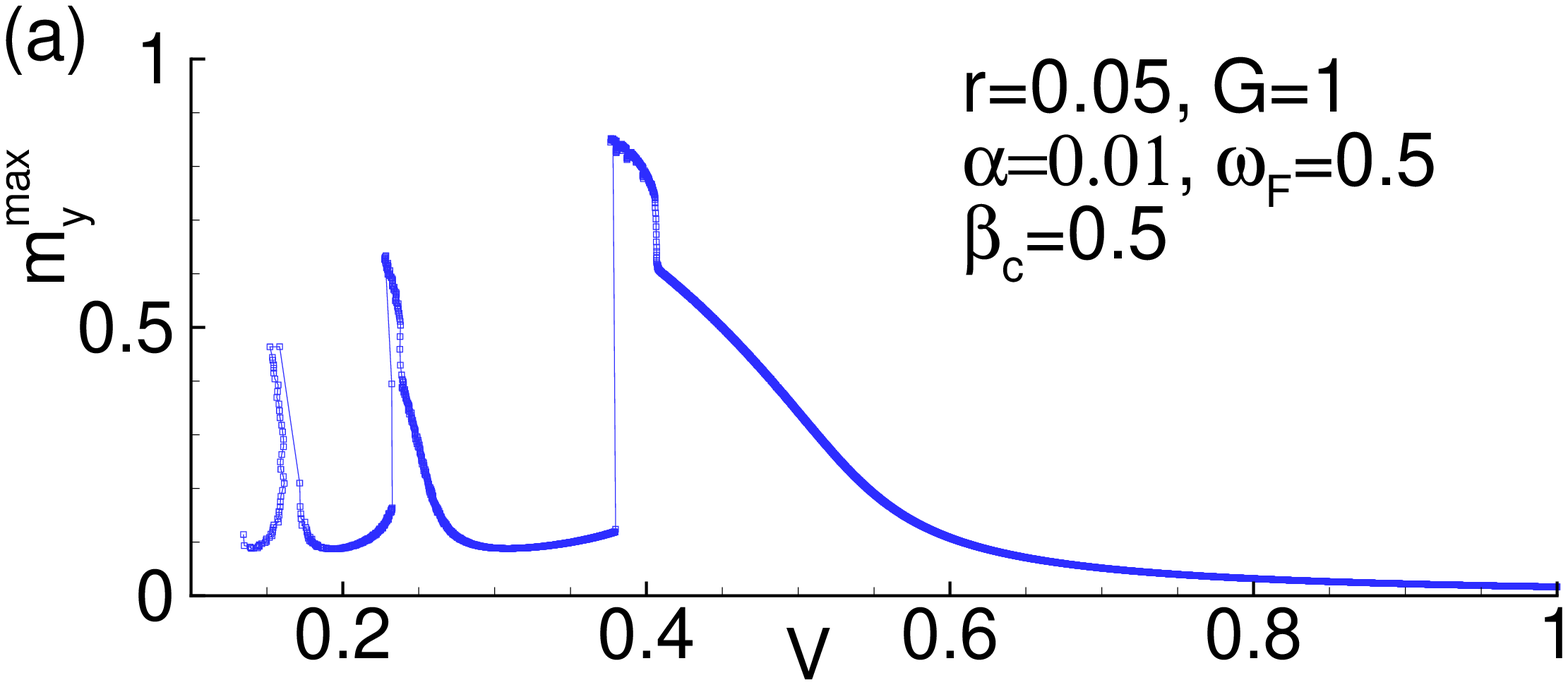}
\includegraphics[width=8cm]{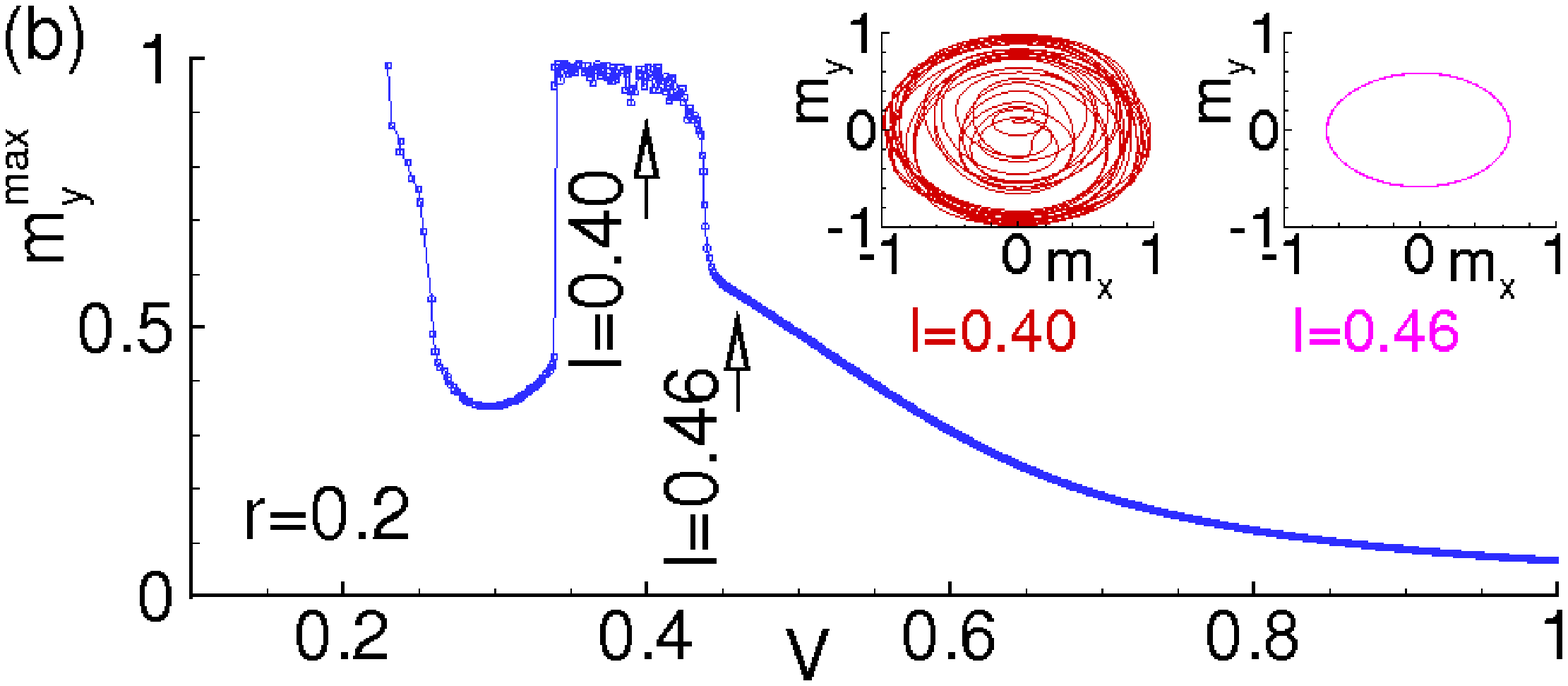}
\includegraphics[width=8cm]{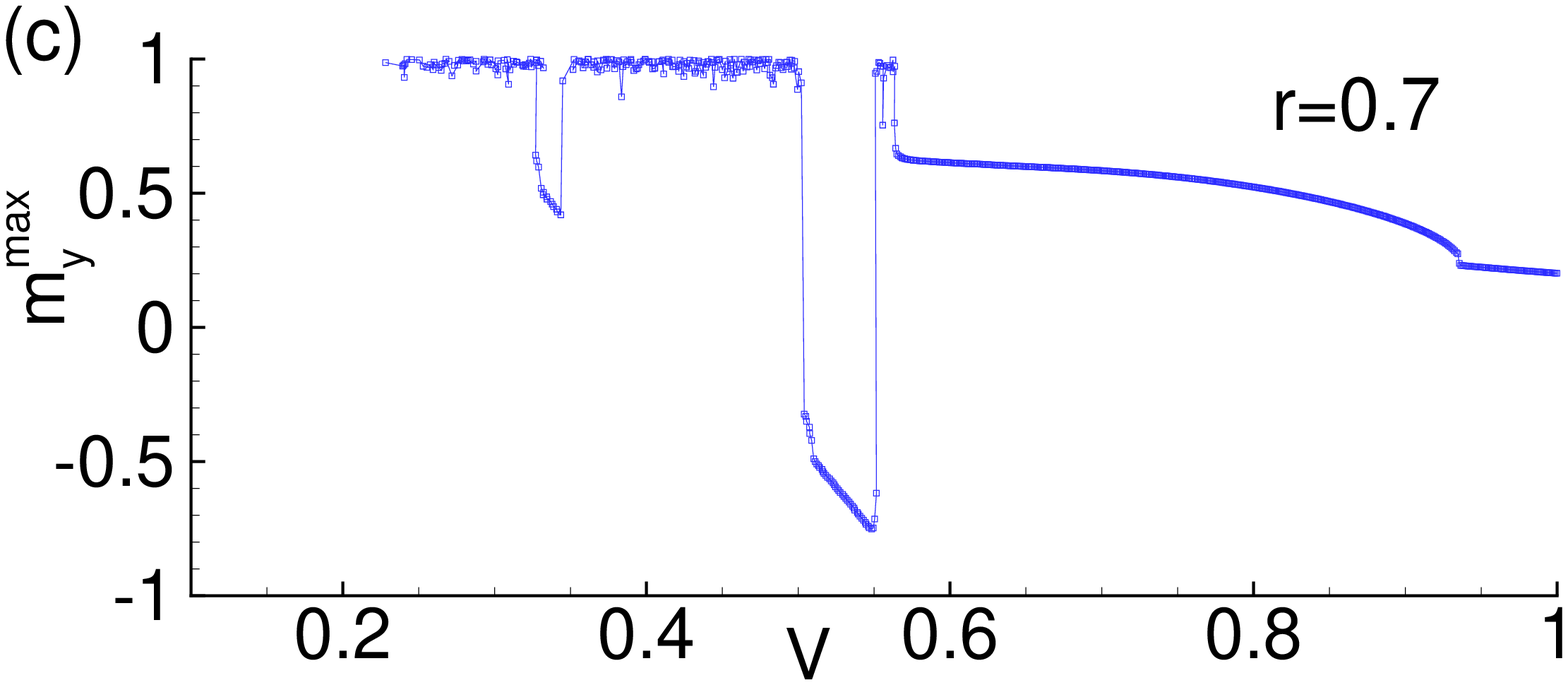}
\includegraphics[width=8cm]{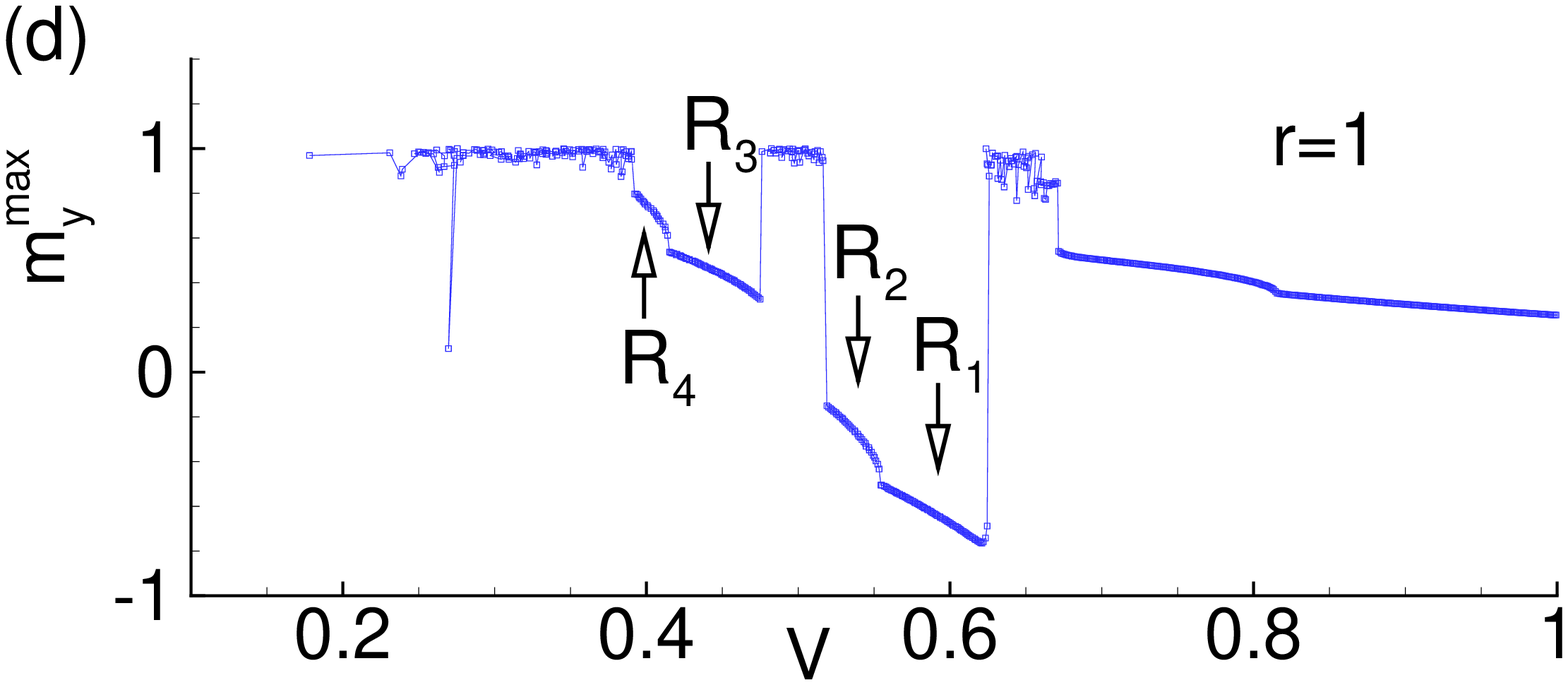}
\caption{Transformation of the ferromagnetic resonance region by changing the spin-orbit coupling (r). Other parameters are the same and indicated in figure (a).} \label{2}
\end{figure}

\section{ Effect of spin-orbit coupling. }
The main ingredient of the considered model is the spin-orbit interaction in the ferromagnetic layer, so it plays a crucial role in the effects discovered in this paper. Figure~\ref{2} presents a transformation of voltage dependence of $m^{max}_y$ by changing the parameter of spin-orbit coupling $r$. To clarify the effect, we have taken a larger value of Josephson to magnetic energy of relation $G$ in comparison with the results presented in Fig.\ \ref{1}.  With increasing in $r$, the peaks are shifted and widened, reflecting the damped resonance, and  their intensity increases. We observe a complex resonance region with different type of magnetization trajectories along the IV-curve. Particularly, in the insets of Fig.~\ref{2}(b) we demonstrate the transformation of trajectories (in the $xy$ plane) with decrease in bias current at $I=0.46$ and $I=0.40$. In Fig.~\ref{2}(d) we can clearly distinguish the regular regions ($R_{1}$, $R_{2}$, $R_{3}$ and $R_{4}$) indicated by arrows. One of the main questions studied in this paper concerns the reaction of the magnetic system to superconducting current reflected in the appearance of such regular regions. So in what follows, we shall
concentrate on the current intervals with regular dynamics.

\begin{figure}[h!]
\centering
\includegraphics[width=8cm]{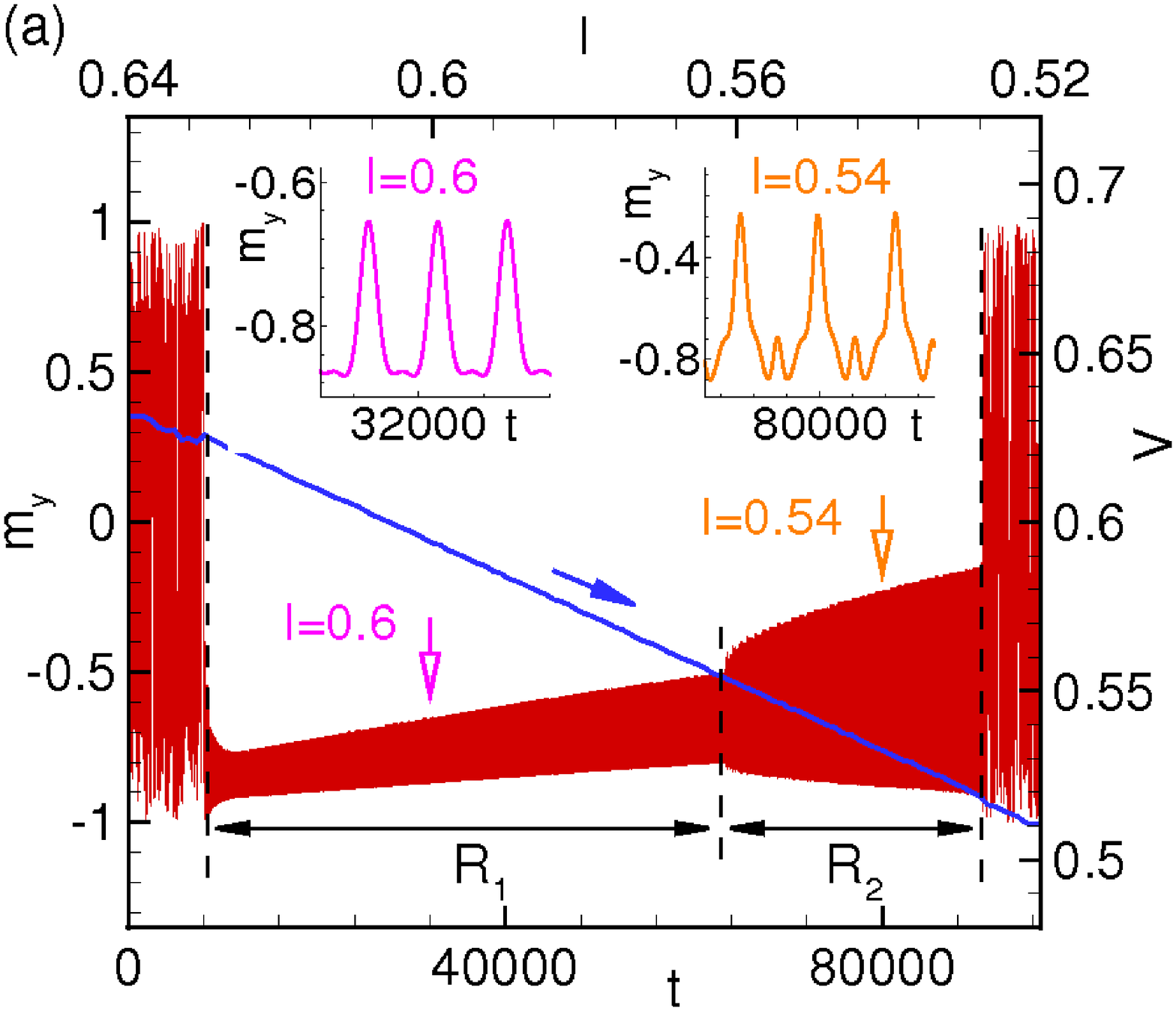}
\includegraphics[width=8cm]{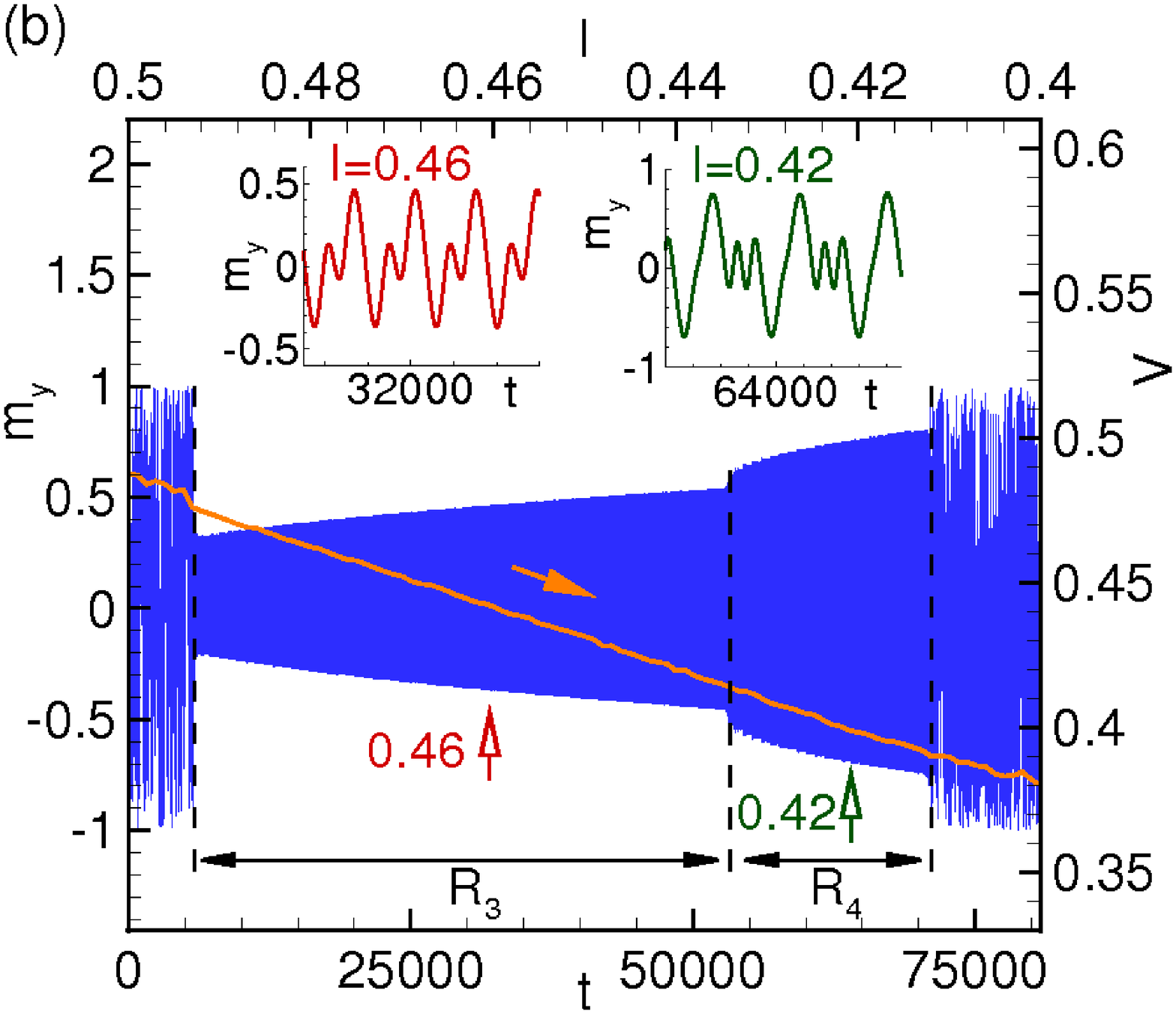}
\caption{(a)  Time dependence of $m_{y}$  in the
regular region shown by arrow in Fig.~\ref{2}(d). Insets demonstrate the character of the $m_y(t)$ oscillations in the $R_1$ and $R_2$ current intervals;  (b) The same for the second region. The lines show the corresponding parts of
IV-characteristics. } \label{3}
\end{figure}

To clarify the dynamics in the selected regular regions, we have investigated the time dependence of the magnetization component $m_y$ presented in Fig.~\ref{3} together with the corresponding parts of  IV-characteristic. Two different parts are clearly pronounced in these both regions where the amplitude of $m_{y}$ grows  with decreasing bias current and has a jump between them. These parts are denoted as $R_1$ and $R_2$ in the first region and $R_3$ and $R_4$ in the second one. Demonstrated in the insets the enlarged time dependence of $m_{y}$ taken at arbitrary currents shows a different character of oscillations. This fact explains the observed jumps between  $R_1$ and $R_2$ in the first region and $R_3$ and $R_4$ in the second one. As we demonstrate below, the origin of different time dependence in these regions is related to the change of the character of magnetic precessions.

\section{ Specific phase trajectories. }
\begin{figure}[h!]
\centering
\includegraphics[width=2.65cm]{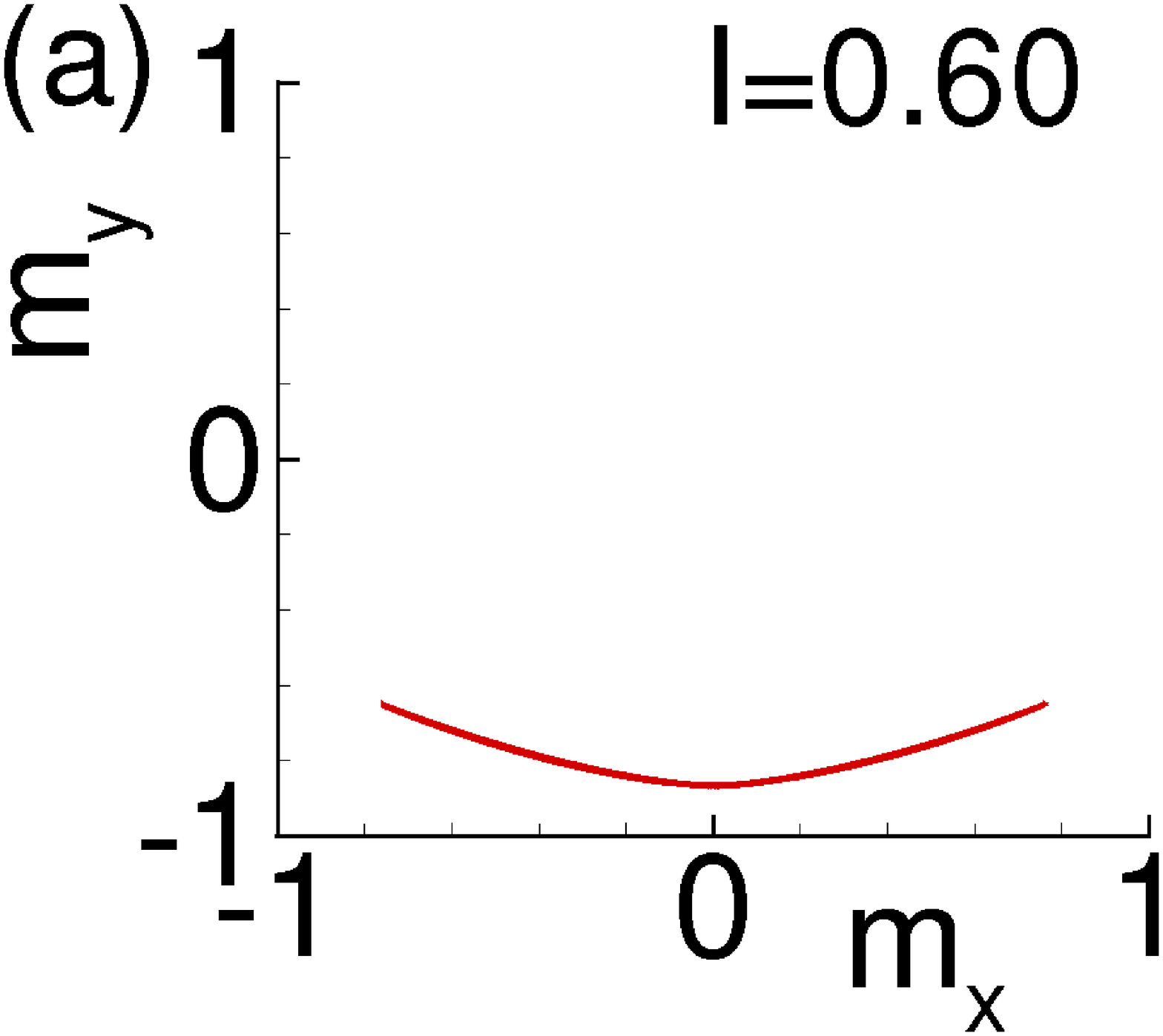}\includegraphics[width=2.65cm]{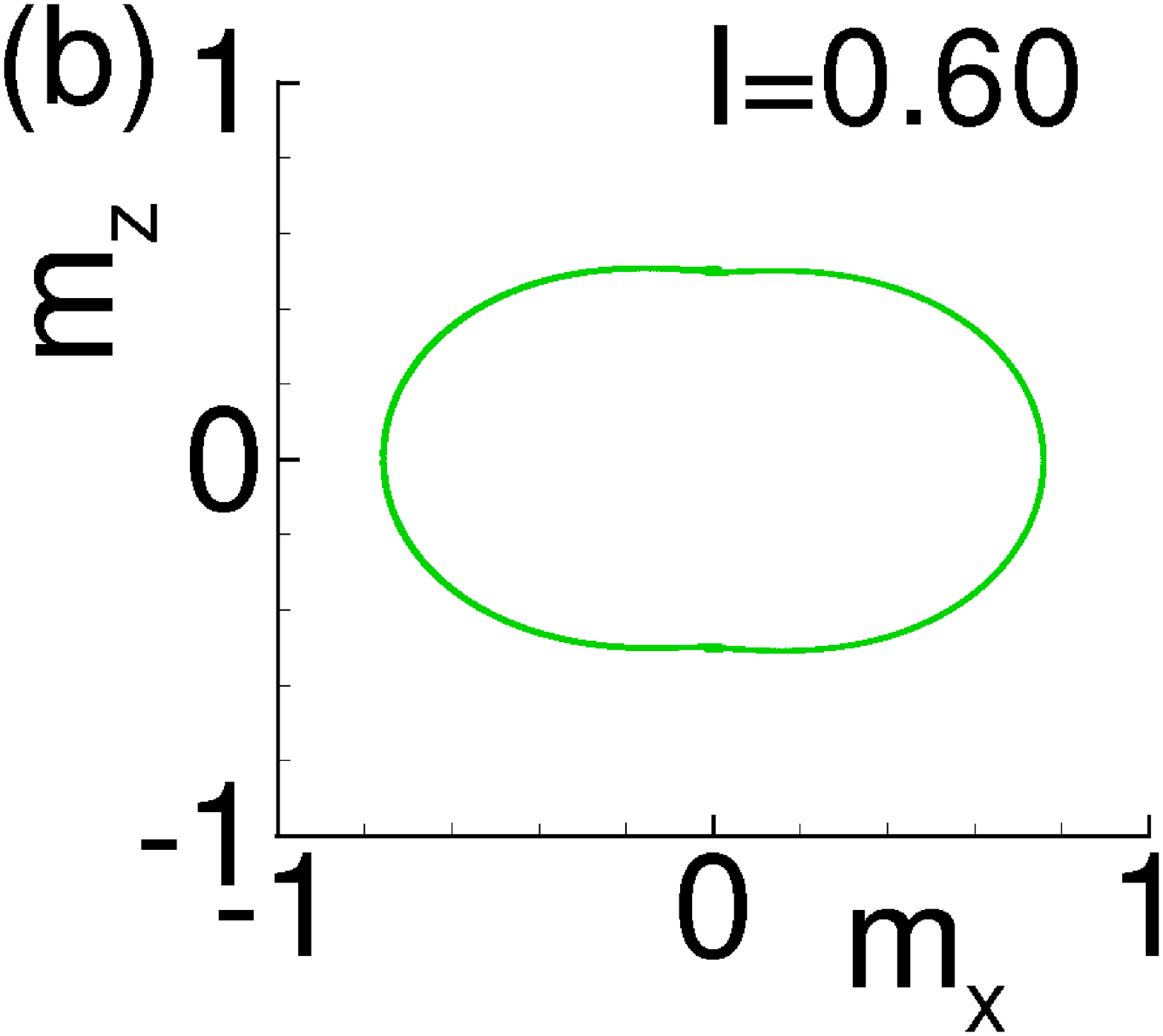}\includegraphics[width=2.65cm]{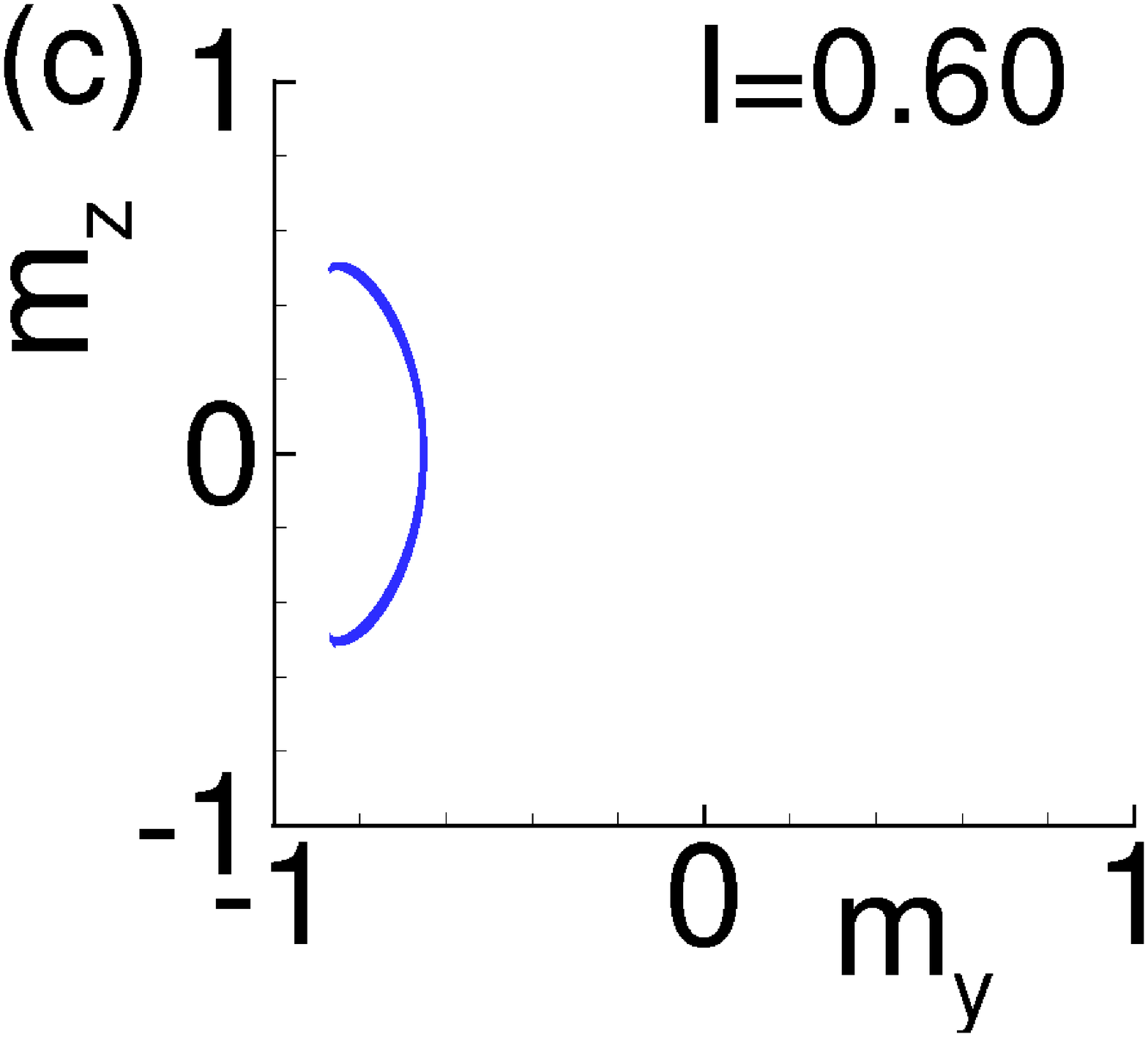}
\includegraphics[width=2.65cm]{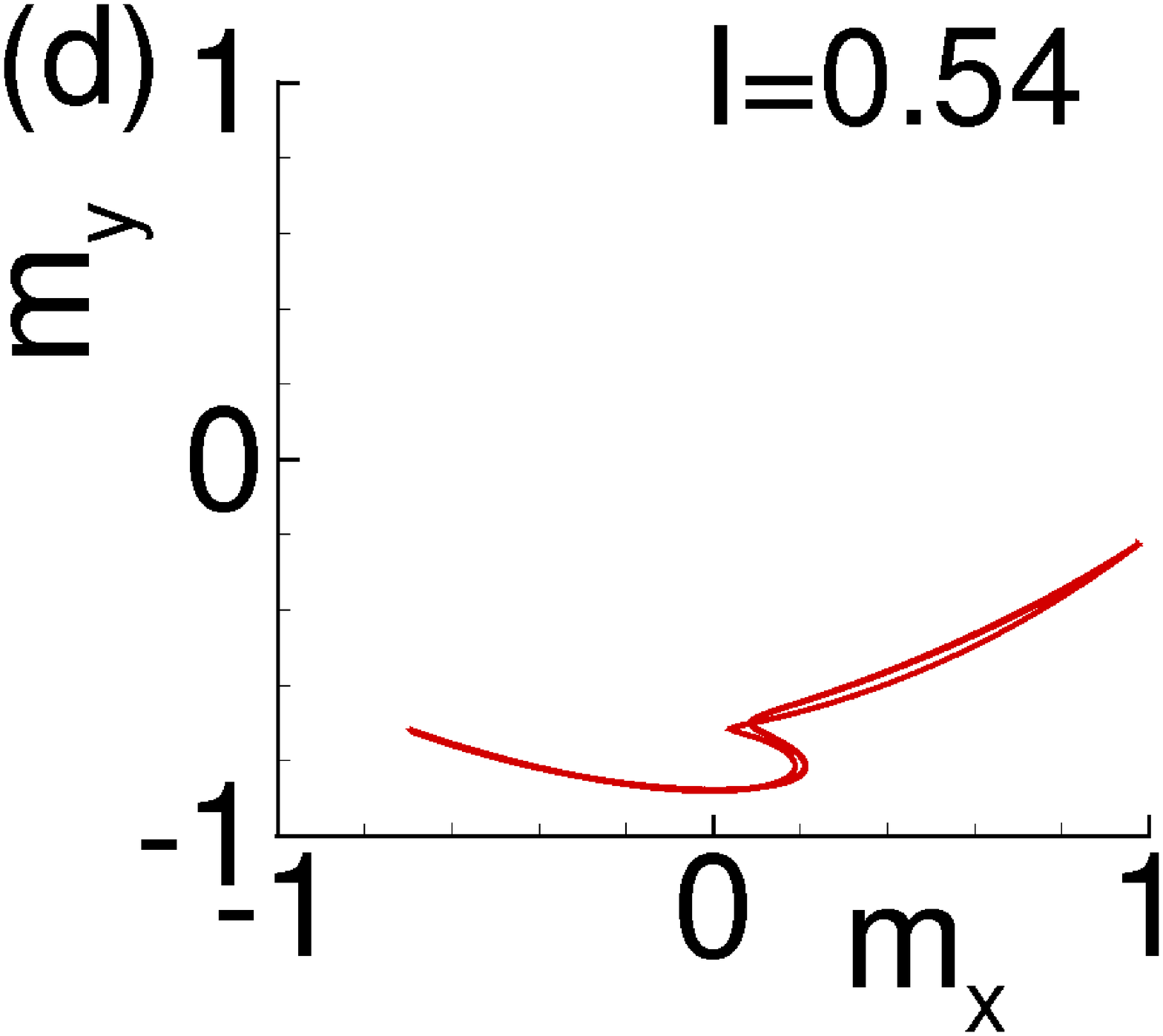}\includegraphics[width=2.65cm]{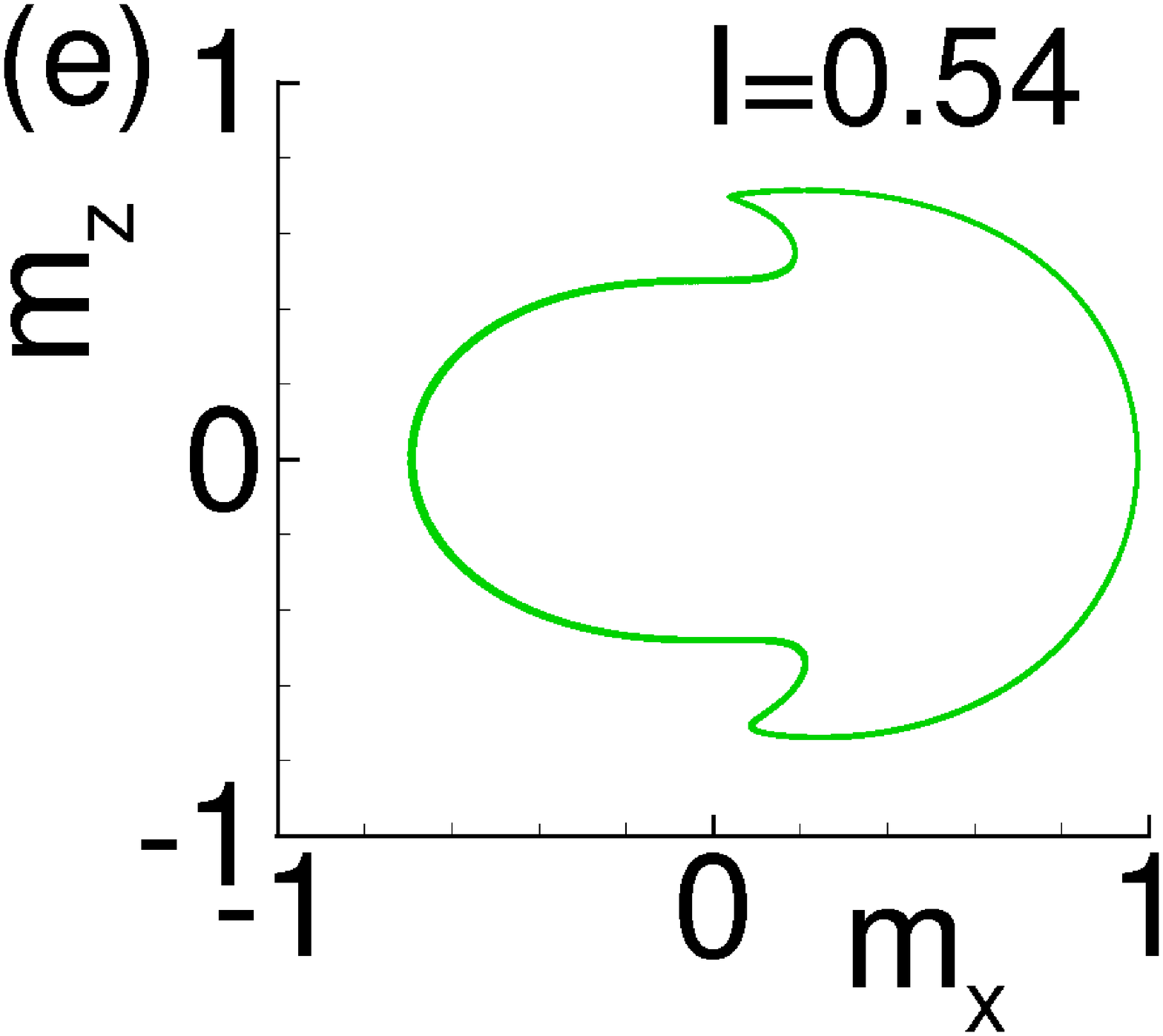}\includegraphics[width=2.65cm]{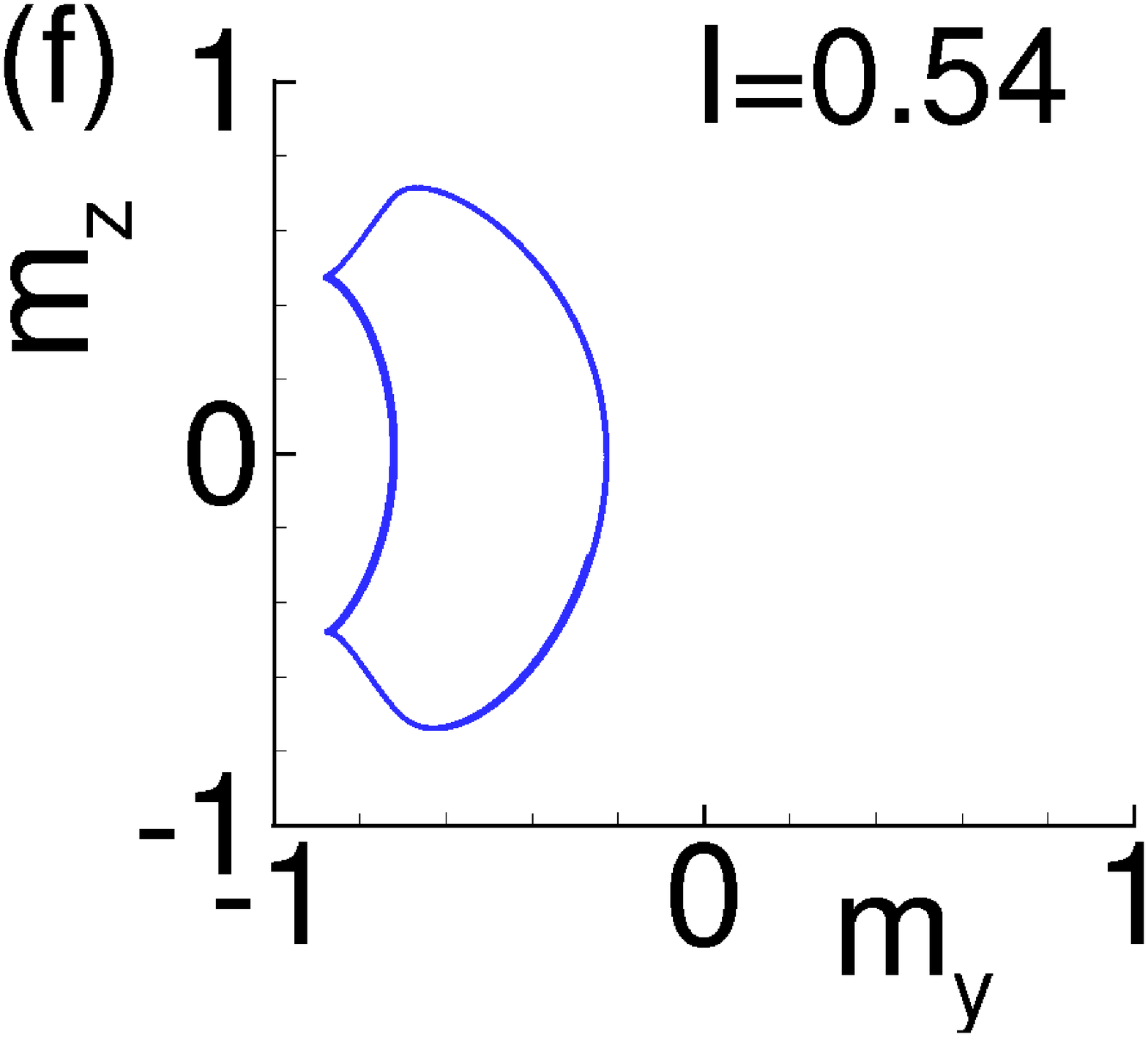}
\includegraphics[width=2.65cm]{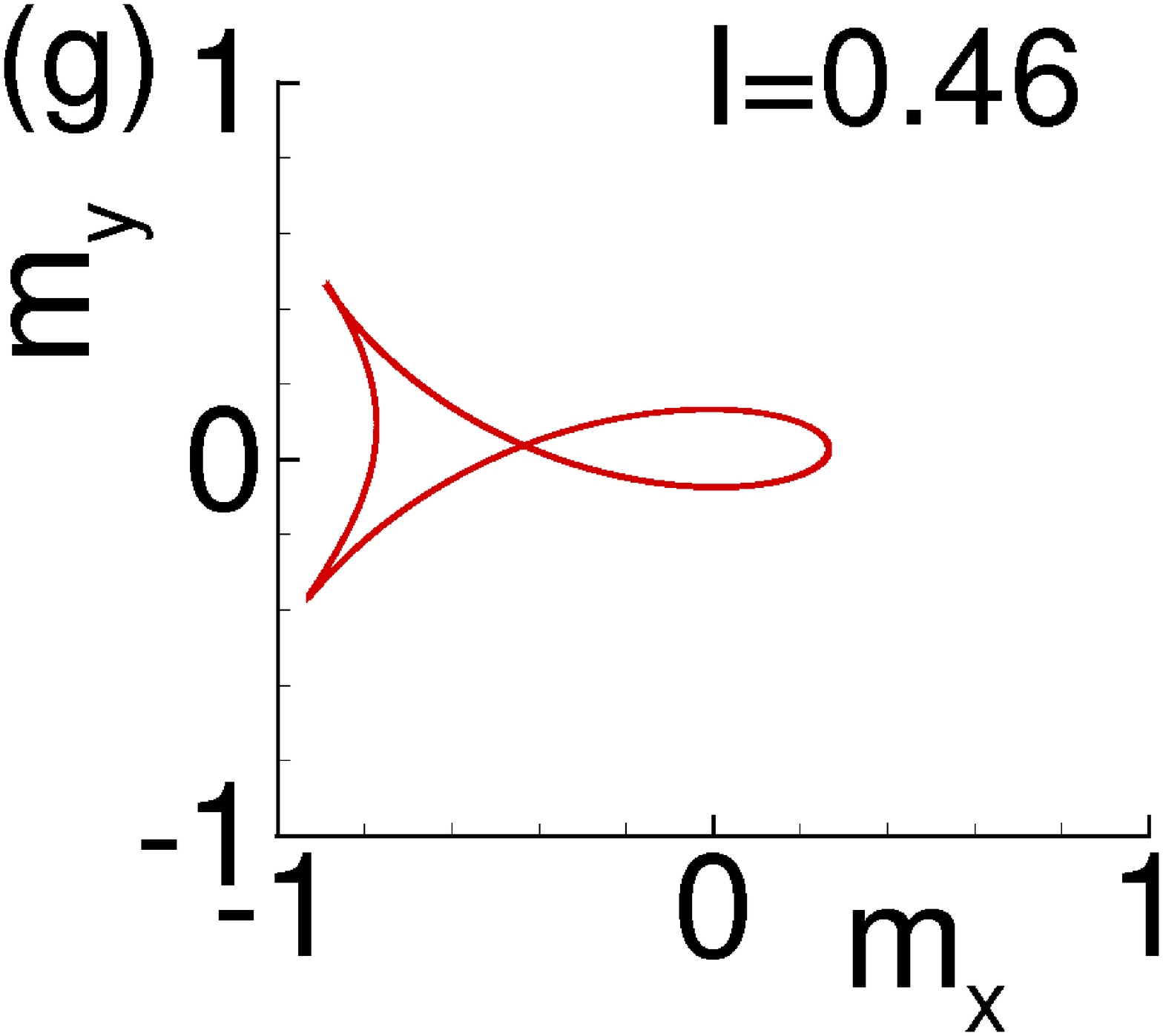}\includegraphics[width=2.65cm]{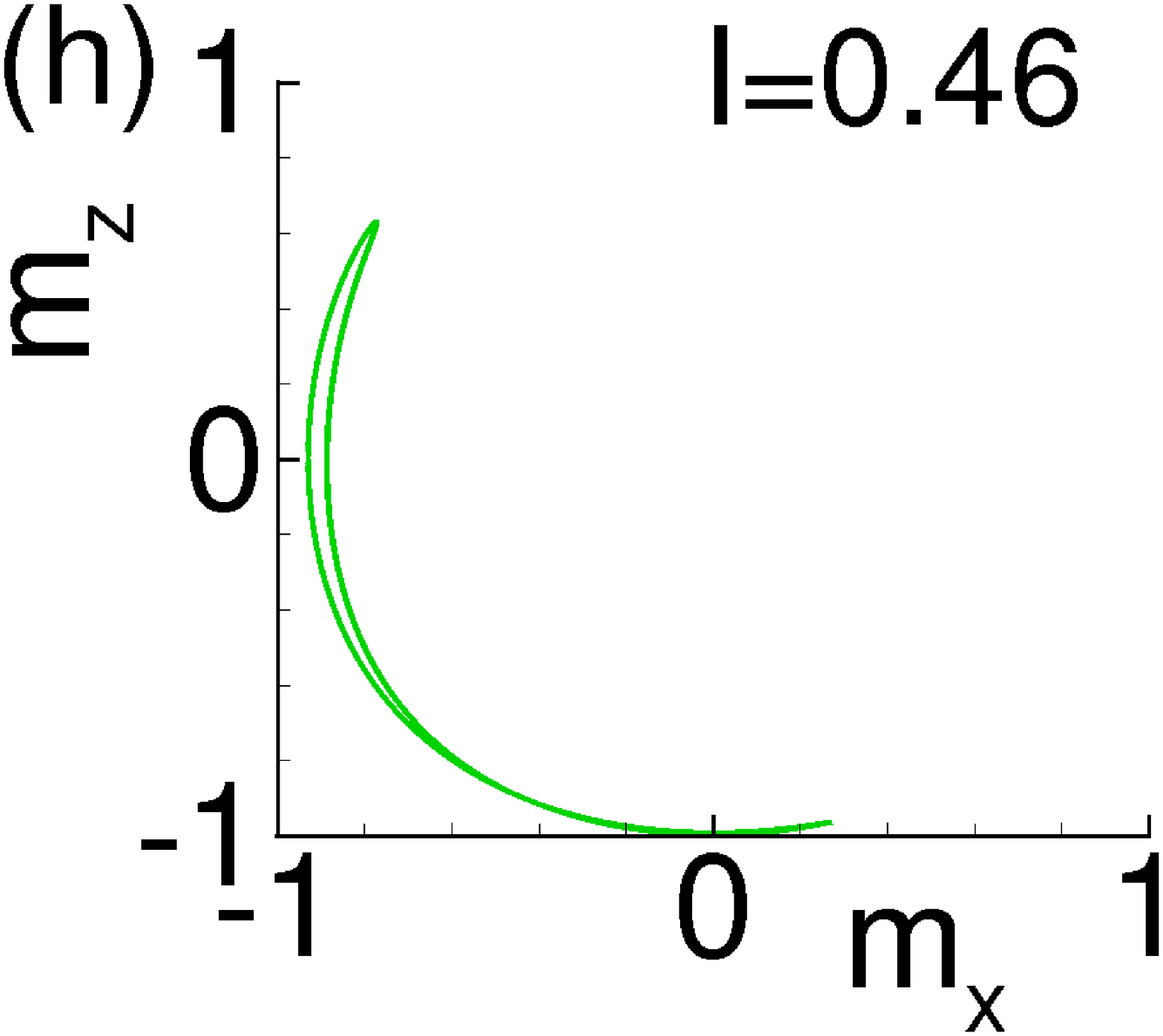}\includegraphics[width=2.65cm]{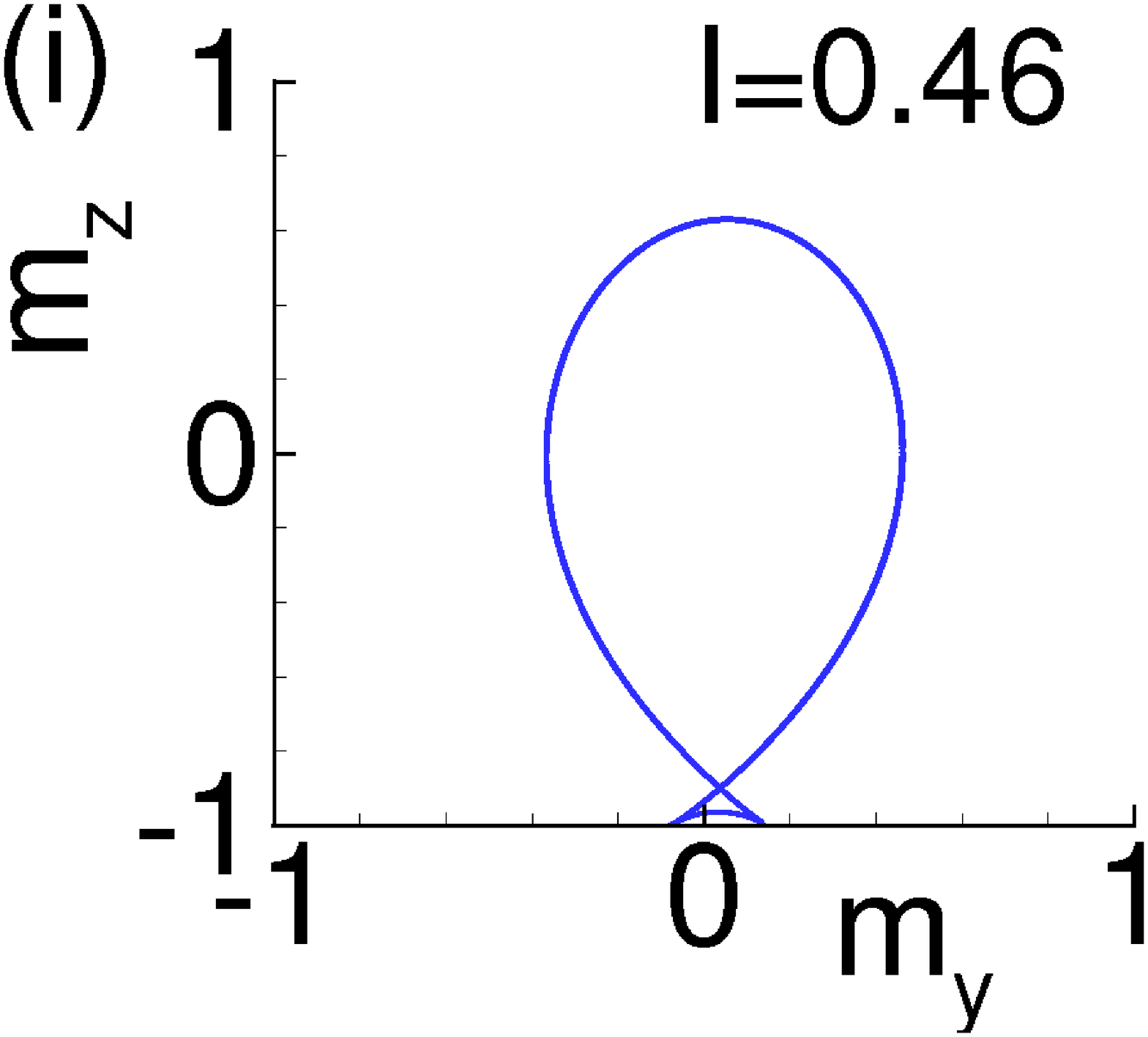}
\includegraphics[width=2.65cm]{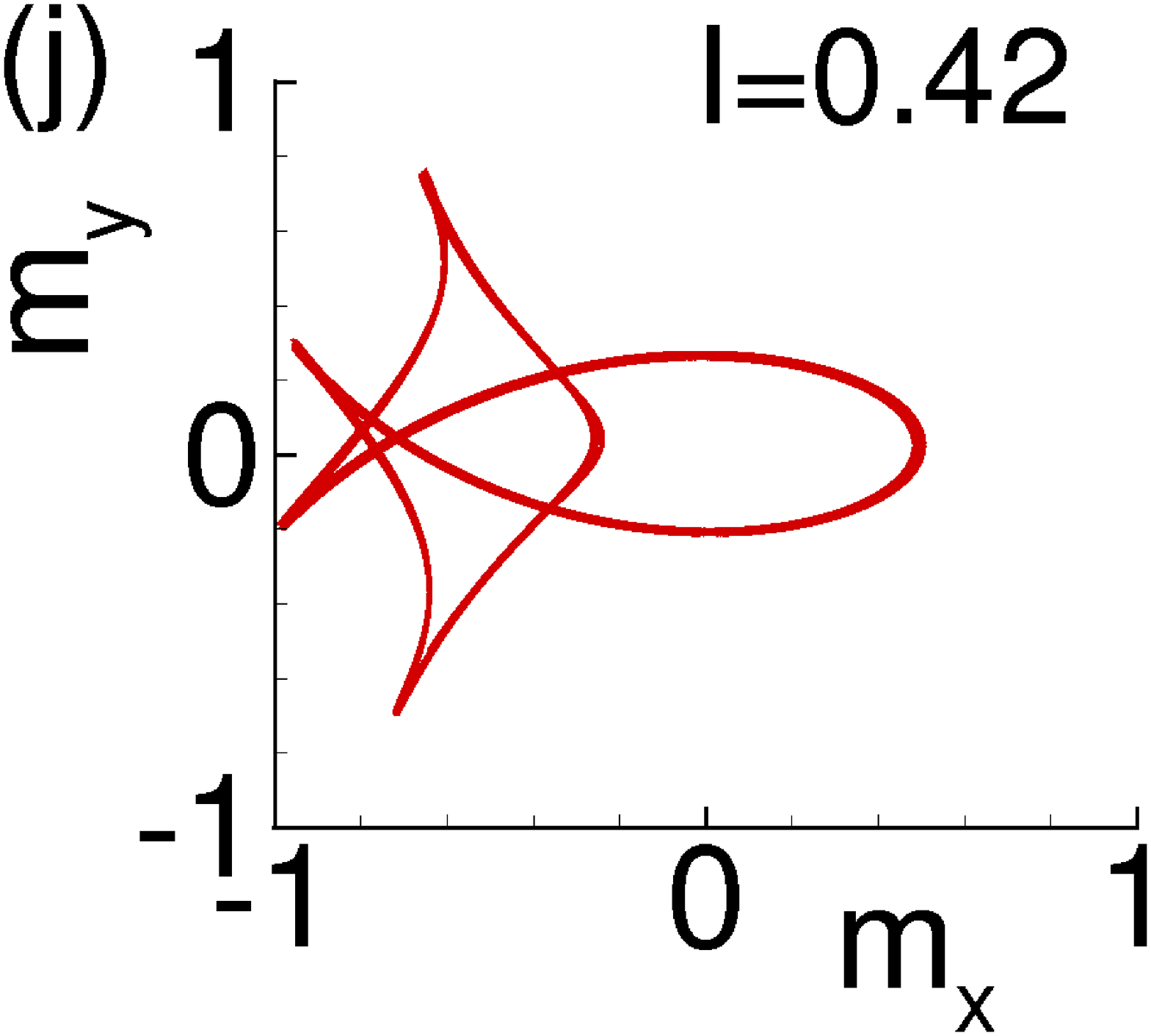}\includegraphics[width=2.65cm]{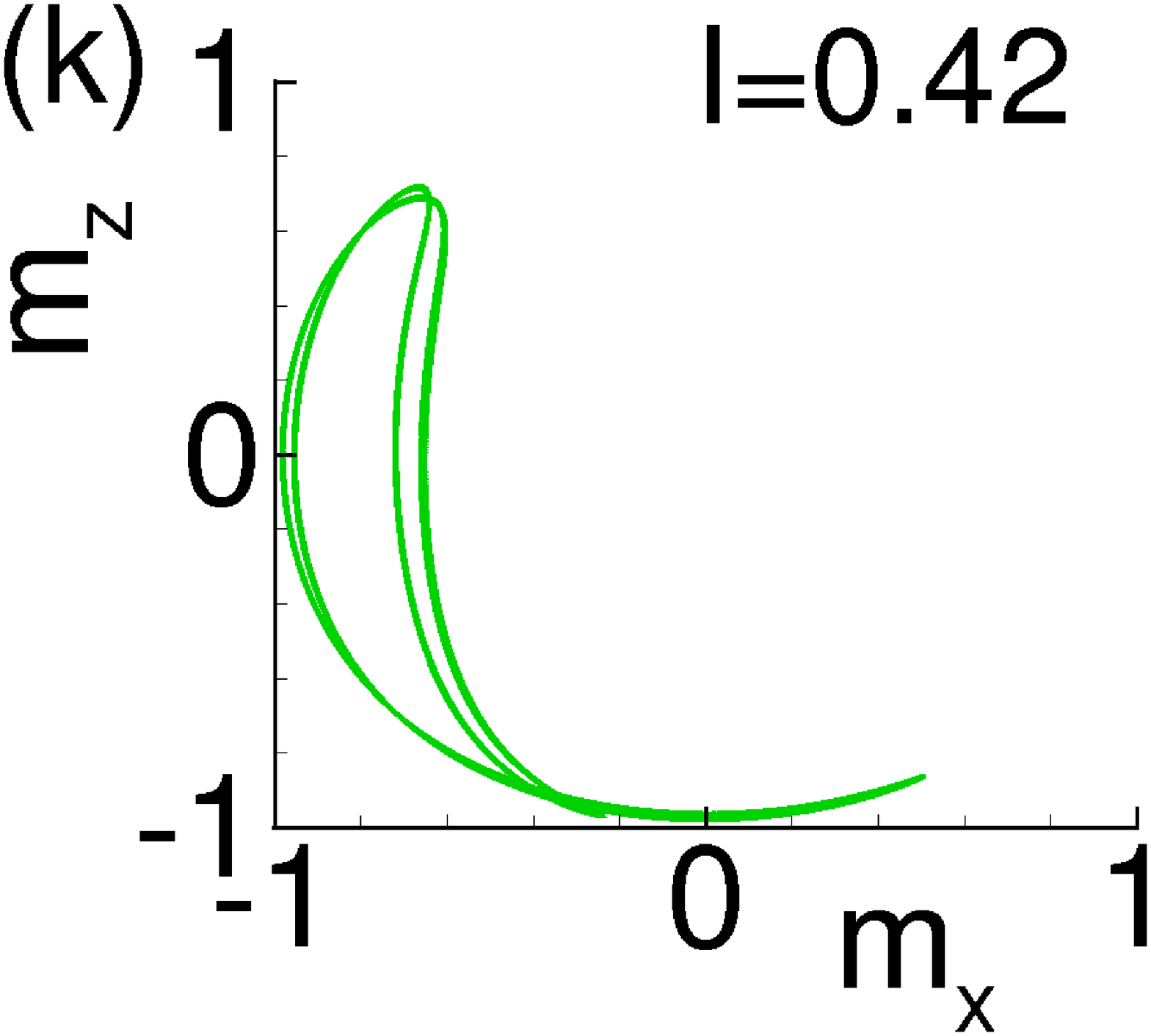}\includegraphics[width=2.65cm]{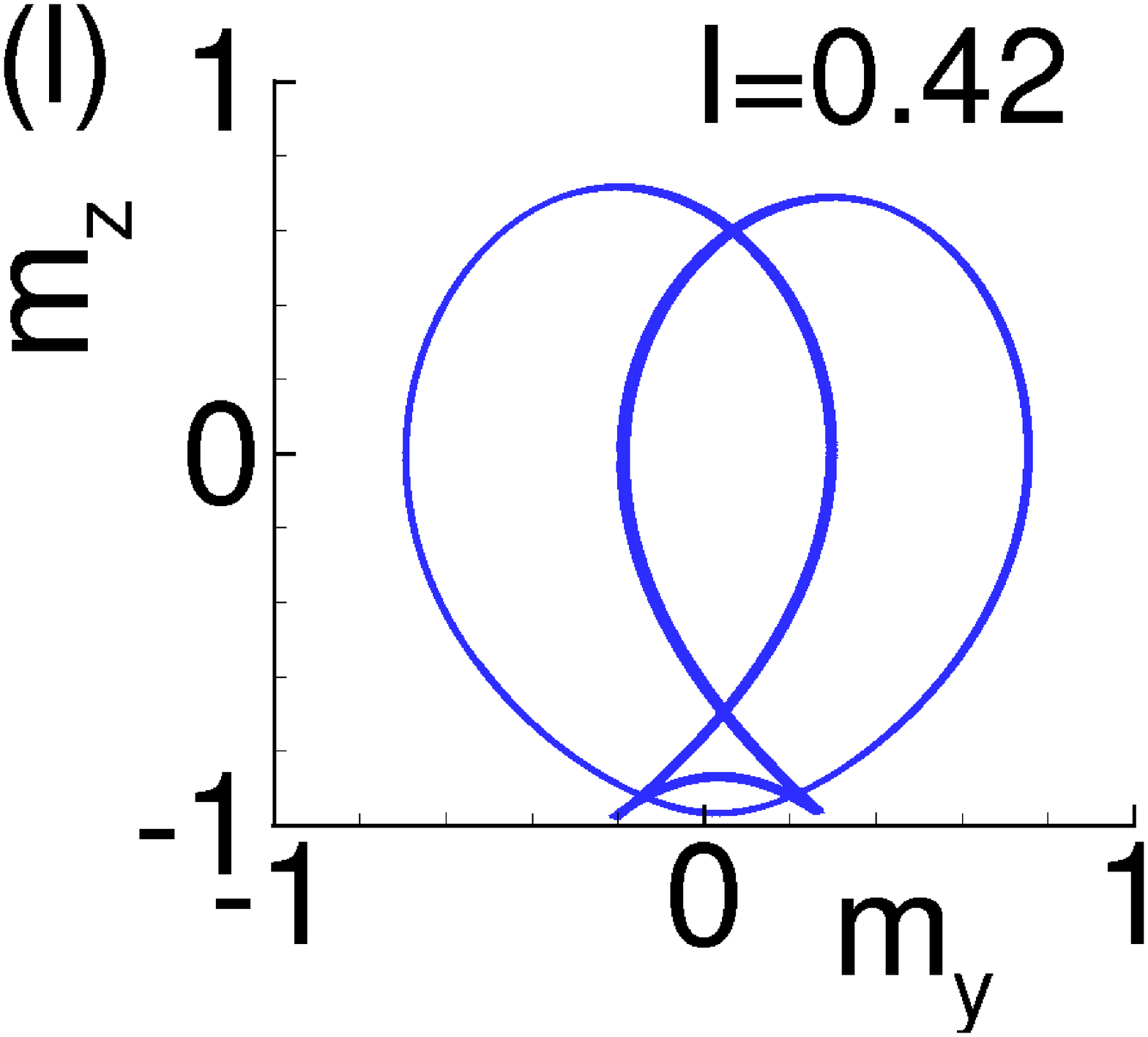}
\caption{Magnetization trajectories in the
 planes $m_{y}$ -- $m_{x}$, $m_{z}$ -- $m_{x}$, and $m_{z}$ -- $m_{y}$ for regular regions $R_i$}
\label{4}
\end{figure}

Here we demonstrate that current intervals $R_i$ indicated in Fig.~\ref{2} and Fig.~\ref{3} are characterized by different  trajectories of the magnetic moment. The characteristic trajectories in
the planes $m_{y}$ -- $m_{x}$, $m_{z}$ -- $m_{x}$, and $m_{z}$ --$m_{y}$ realized in these regions are shown in Fig.\ref{4} for four values of bias current $I=0.60$, $0.54$, $0.46$ and $0.42$.
We see different specific forms of trajectories and some of them for
distinctness we call ``apple'' (b), ``sickle'' (d)
``mushroom'' (e), ``fish'' (g) and  ``moon'' (h)). The first current interval $R_1$ is
characterized by ``apple'' type dynamics demonstrated in Fig.\ref{4}(b) at
$I=0.60$. With decreasing bias current we observe a transformation of trajectories of
``apple'' to ``mushroom'' type in the $m_{z}-m_{x}$ plane, while the
third interval $R_3$ is characterized by
``fish'' and ``moon'' type trajectories. In the fourth interval $R_4$ we observe the characteristic trajectory of ``double fish'' type. \emph{Thus, the presented results demonstrate a unique possibility of
controlling the magnetization dynamics via external bias current.
}
\begin{figure}[h!]
\centering
\includegraphics[width=4cm]{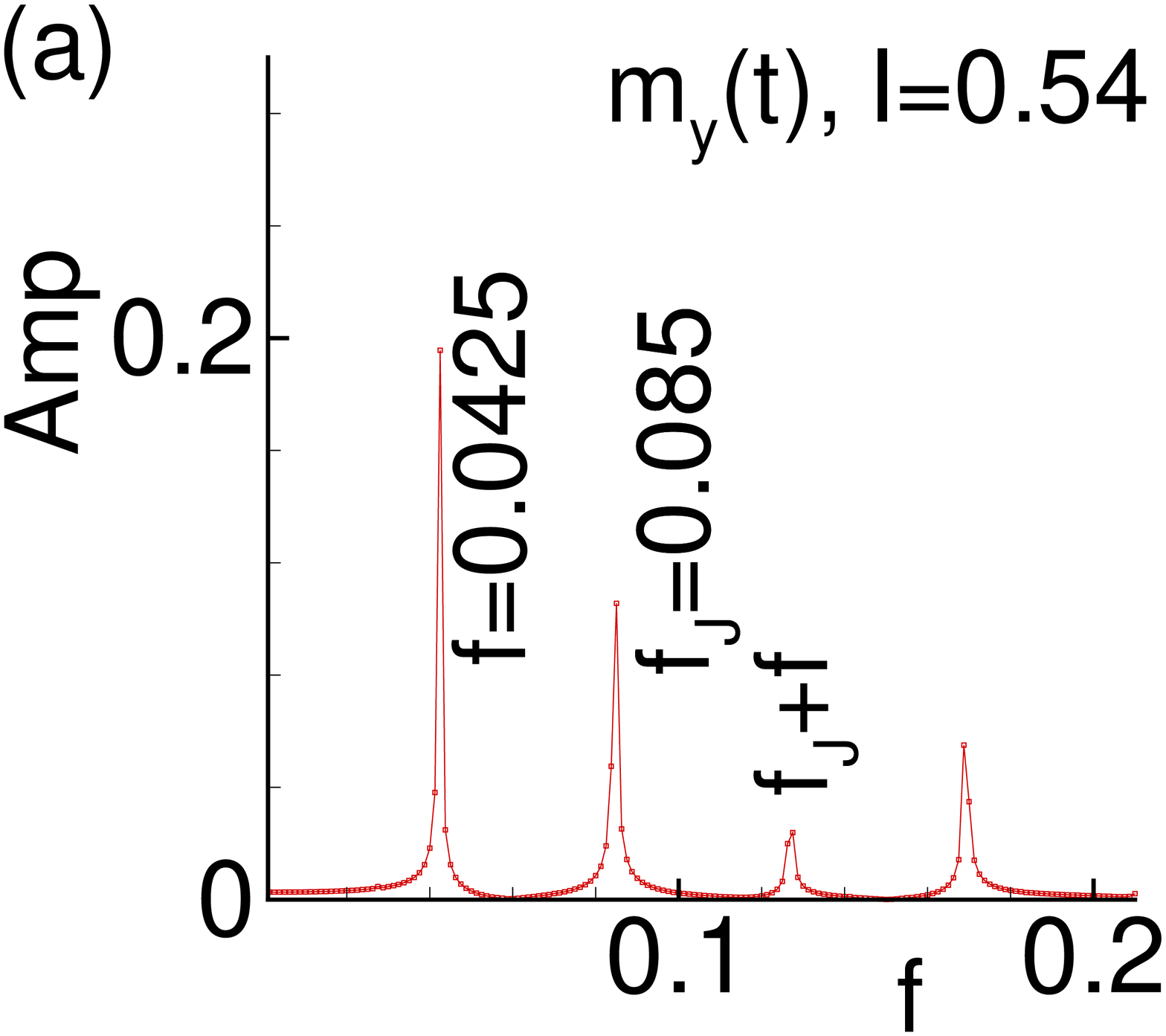}\includegraphics[width=4cm]{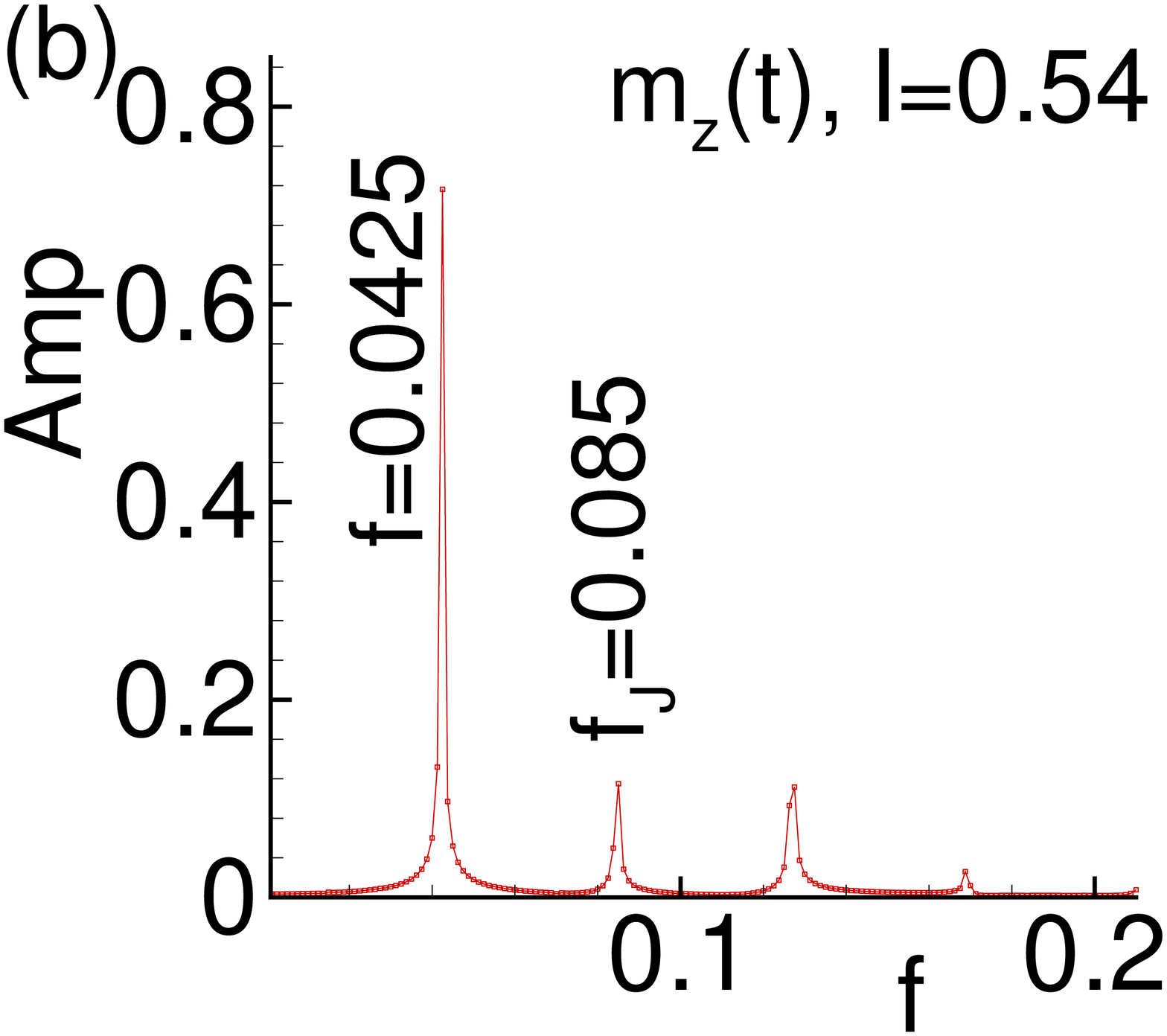}
\includegraphics[width=4cm]{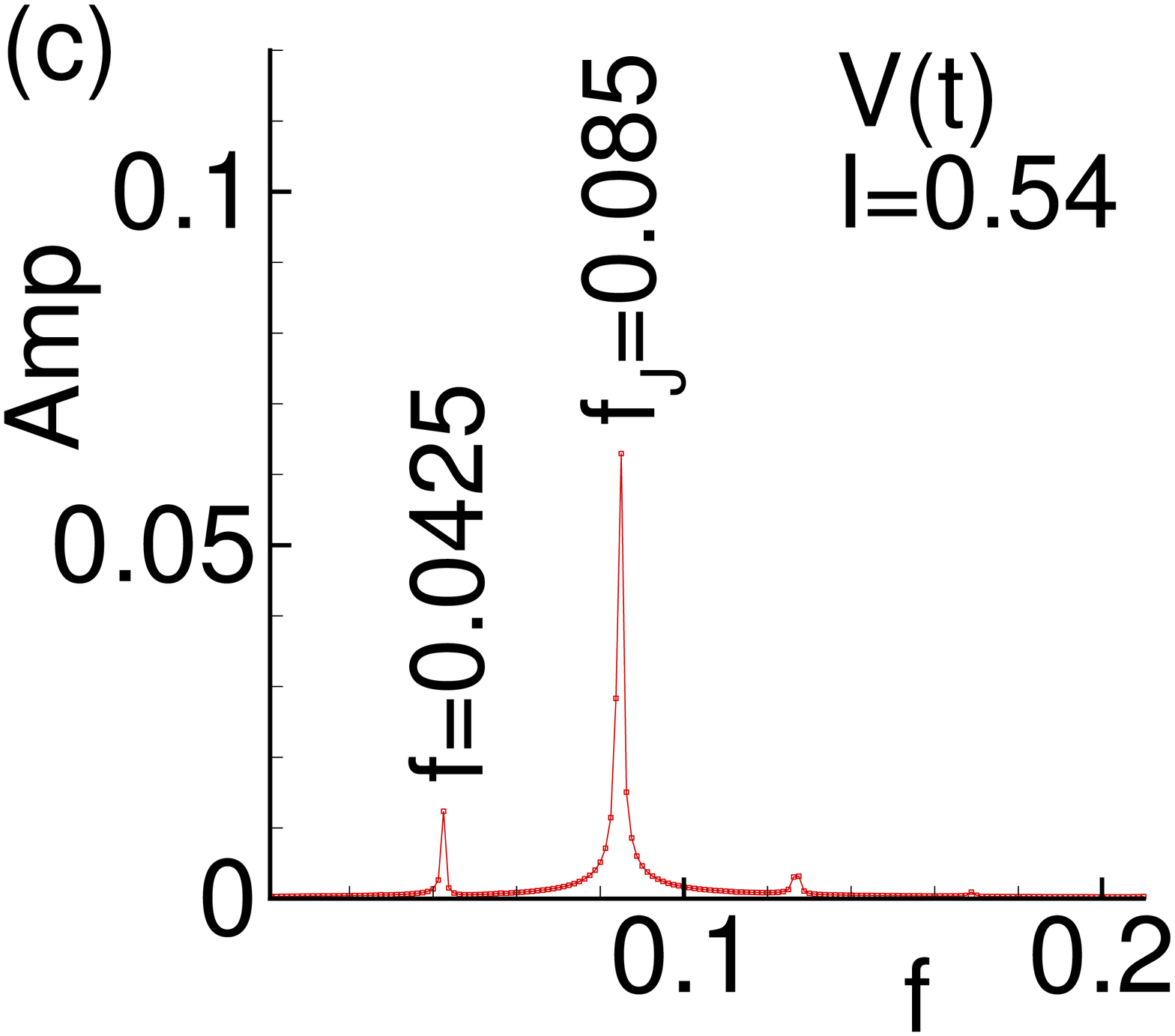}\includegraphics[width=4cm]{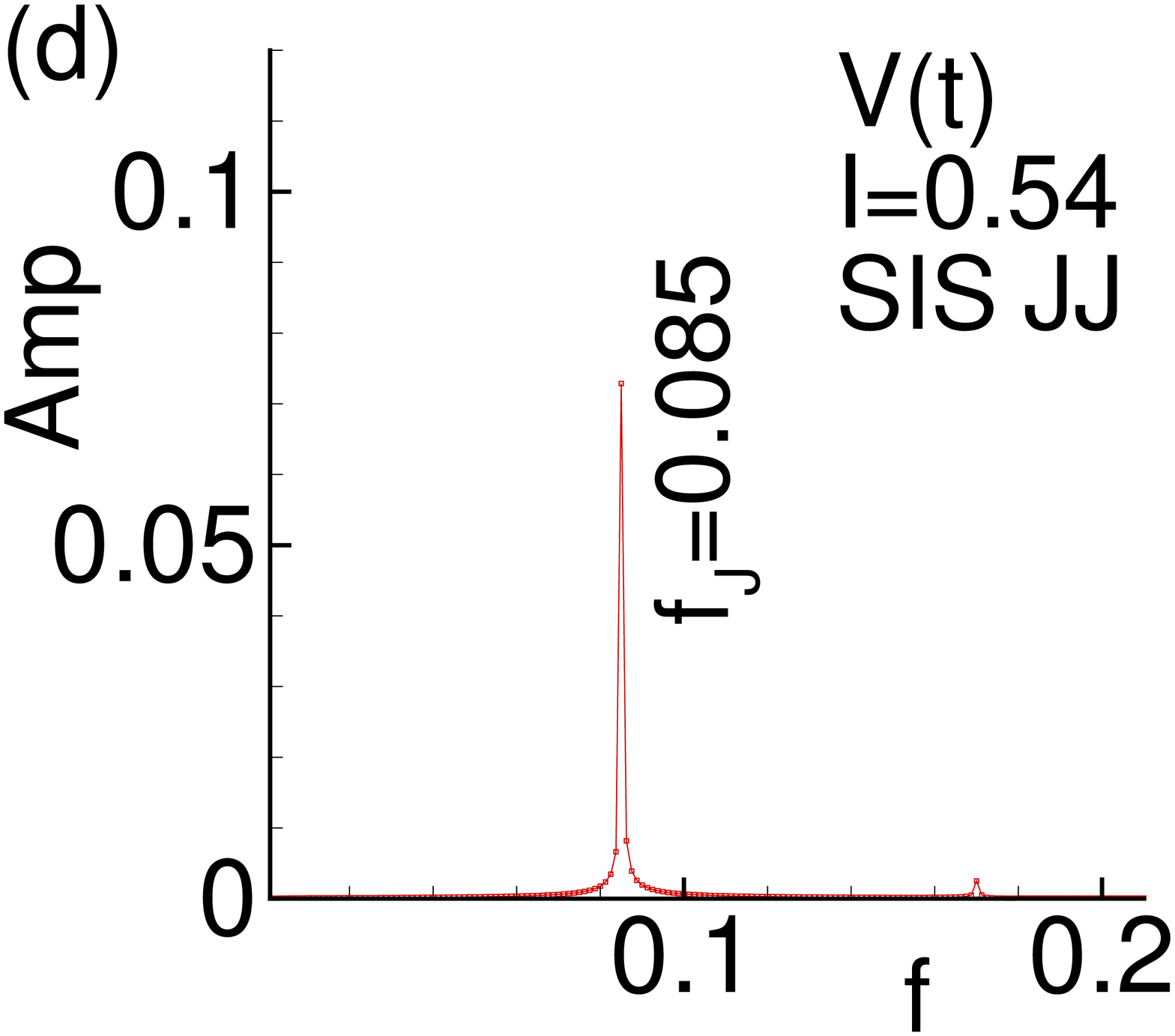}
\caption{FFT analysis of time dependencies of: (a) $m_{y}(t)$; (b)
$m_{z}(t)$; (c) $V(t)$; (d)$V(t)$ for JJ without the magnetic system at
$I=0.54$.} \label{5}
\end{figure}

To test whether the observed temporal dependencies of magnetization are related to the Josephson oscillations, we have made a detailed FFT analysis at different values of bias current.   In particular,  in Fig.\ \ref{5} we present the results of FFT analysis of the time dependence of the magnetization components and
voltage for JJ with and without the magnetic system at $I=0.54$. Comparing the results presented in this figure, we find that the dynamics of magnetization  in this case is really determined by
Josephson frequency $f_J=\omega_{J}/(2\pi)=0.085$. The existence of half harmonics in
this parameter regime indicates that the excitation of magnetic
dynamics happens parametrically. We also note the effect of the
magnetic oscillations on Josephson current which is manifested as a
small peak in FFT of $V(t)$. This peak is absent for the Josephson junction superconductor-insulator-superconductor without the ferromagnetic layer (SIS JJ) as it is demonstrated in Figure~\ref{5}(d). The Results of detailed FFT analysis of dynamics magnetization in the $\varphi_0$ junction at different values of bias current will
be presented somewhere else~\cite{prb-ivc}.

\section{Effects of external radiation}

\begin{figure}[h!]
\centering
\includegraphics[width=4.0cm]{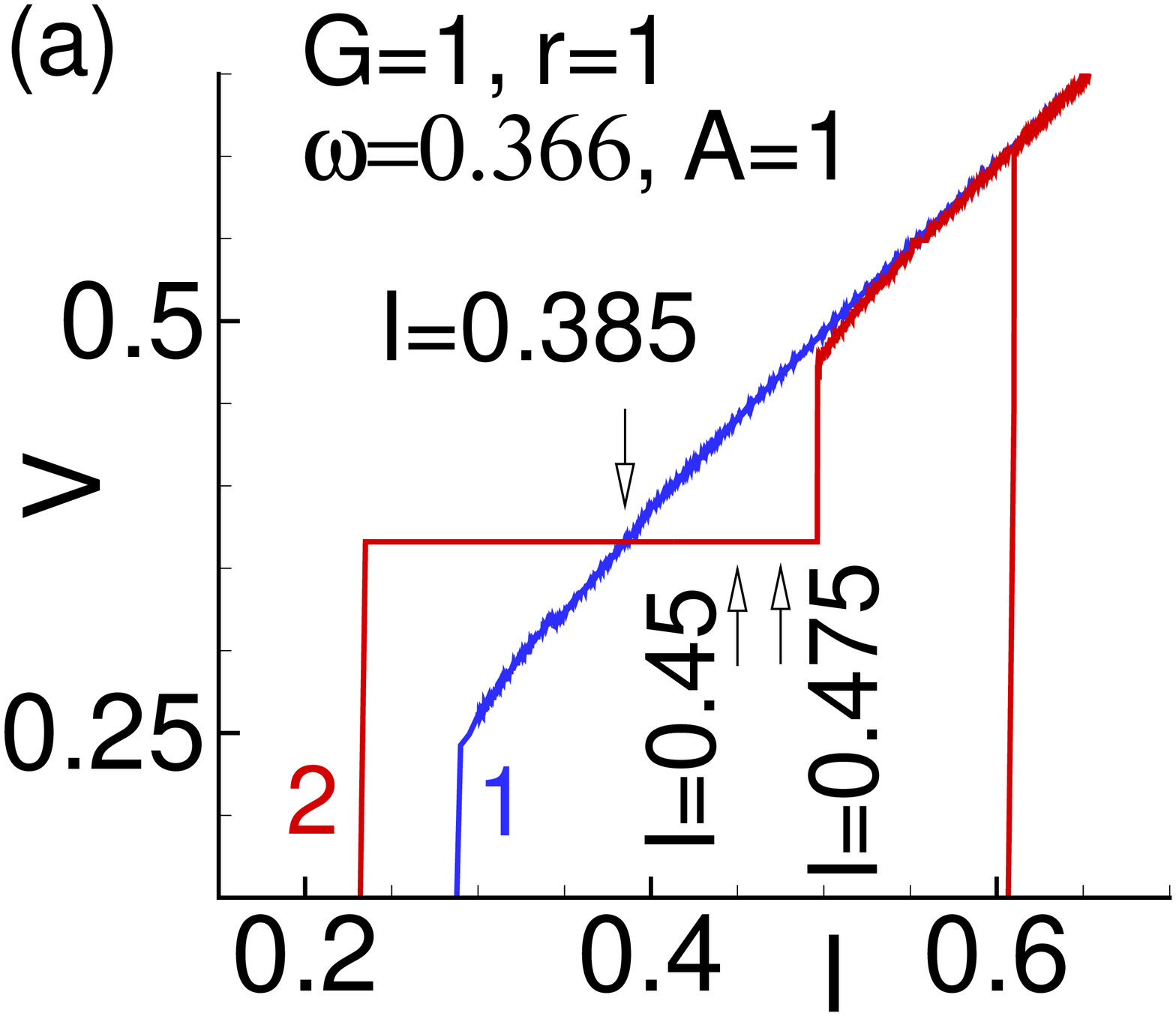}\includegraphics[width=4.0cm]{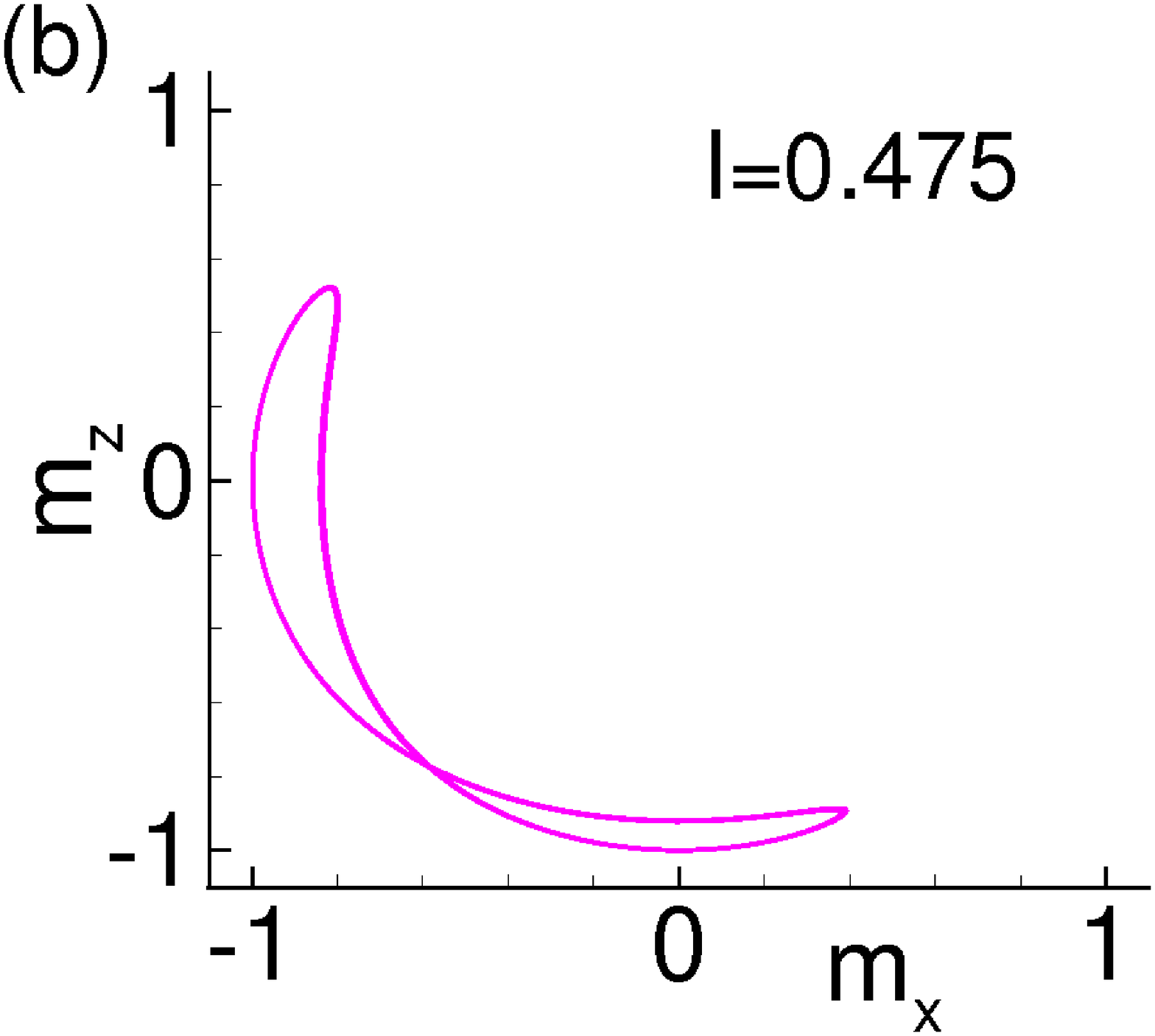}
\includegraphics[width=4.0cm]{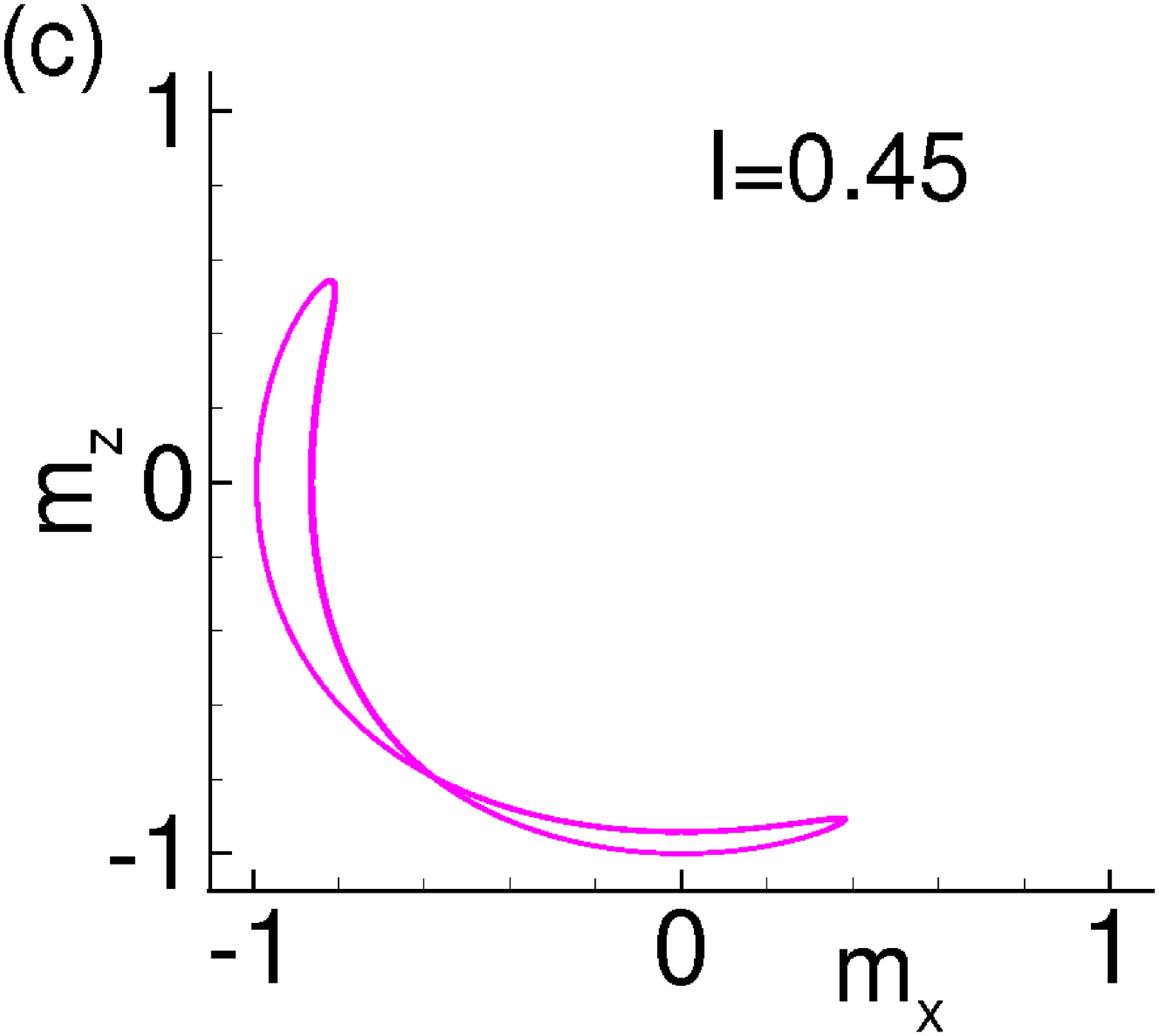}\includegraphics[width=4.0cm]{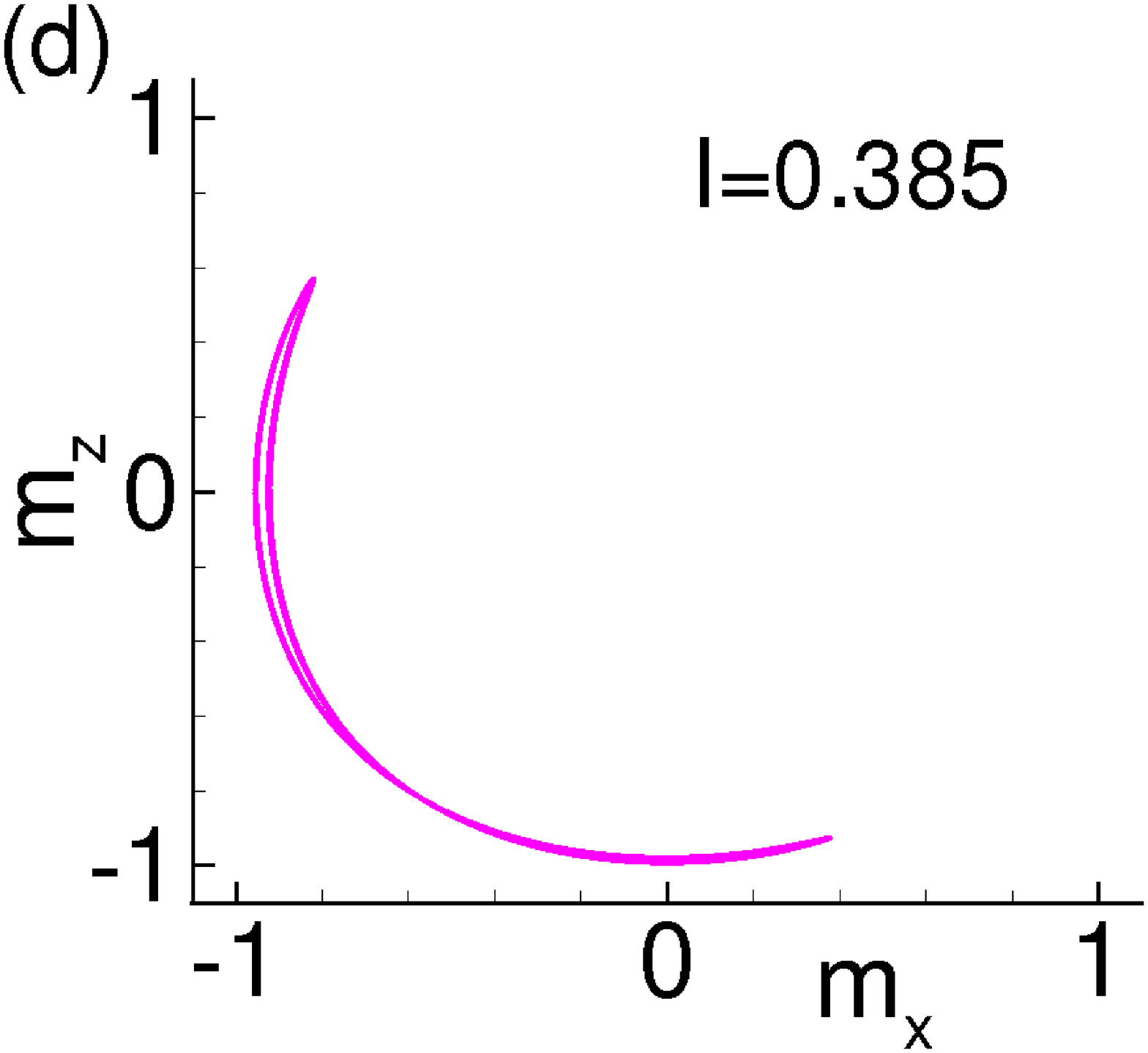}
\caption{(a) IV-characteristics of JJ without (1) and with (2)
radiation. Arrows indicate the bias current values where dynamics of
magnetic precessions was investigated; (b) Dynamics of magnetic
precession in the $m_{z}$ -- $m_{x}$ plane in the presence of external
radiation at $I=0.475$; (c) The same at $I=0.45$; (d) The same at
$I=0.385$.} \label{6}
\end{figure}

Another important feature of the studied magnetization dynamics
concerns the possibility of its control via external electromagnetic
radiation. The presence of such a radiation amounts to $I \to I(t)=
I + A \sin(\omega t)$ in Eq.\ (\ref{syseq}), where $\omega$ is the
frequency and $A$ is the amplitude of the external
radiation \cite{kleiner-book}. We find that such an external
radiation can control the qualitative nature of the magnetic
precession in current interval corresponding to a Shapiro step. To
demonstrate this feature, we show the IV-characteristic of
$\varphi_0$ junction under external radiation with frequency
$\omega=0.366$ and amplitude $A=1$ (which demonstrates the
corresponding Shapiro step at $V=0.366$) in Fig.\ \ref{6}(a). The
resulting magnetization precession in the $m_{z}-m_{x}$ plane at
$I=0.475$, $I=0.45$ and $I=0.385$ are presented in Fig.~\ref{6}(b),
Fig.~\ref{6}(c) and Fig.~\ref{6}(d), respectively. In sharp contrast
to magnetization dynamics without radiation (see Fig.\ \ref{4})
demonstrating different specific precession dynamics with changing
in bias current, the dynamics of magnetic precessions along the
Shapiro step are very similar for all three current values.

\begin{figure}[h!]
\centering
\includegraphics[width=4.0cm]{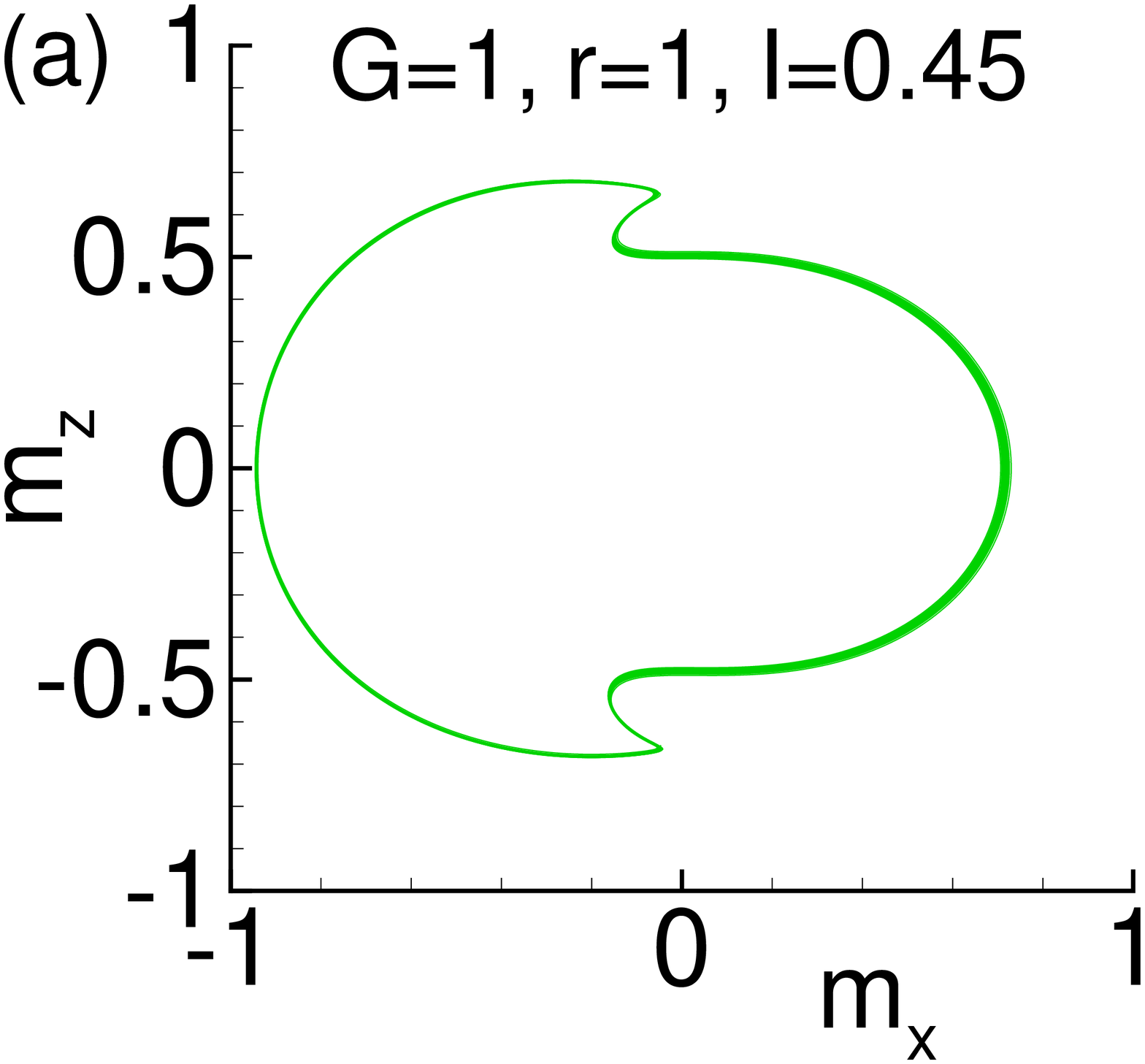}\includegraphics[width=4.0cm]{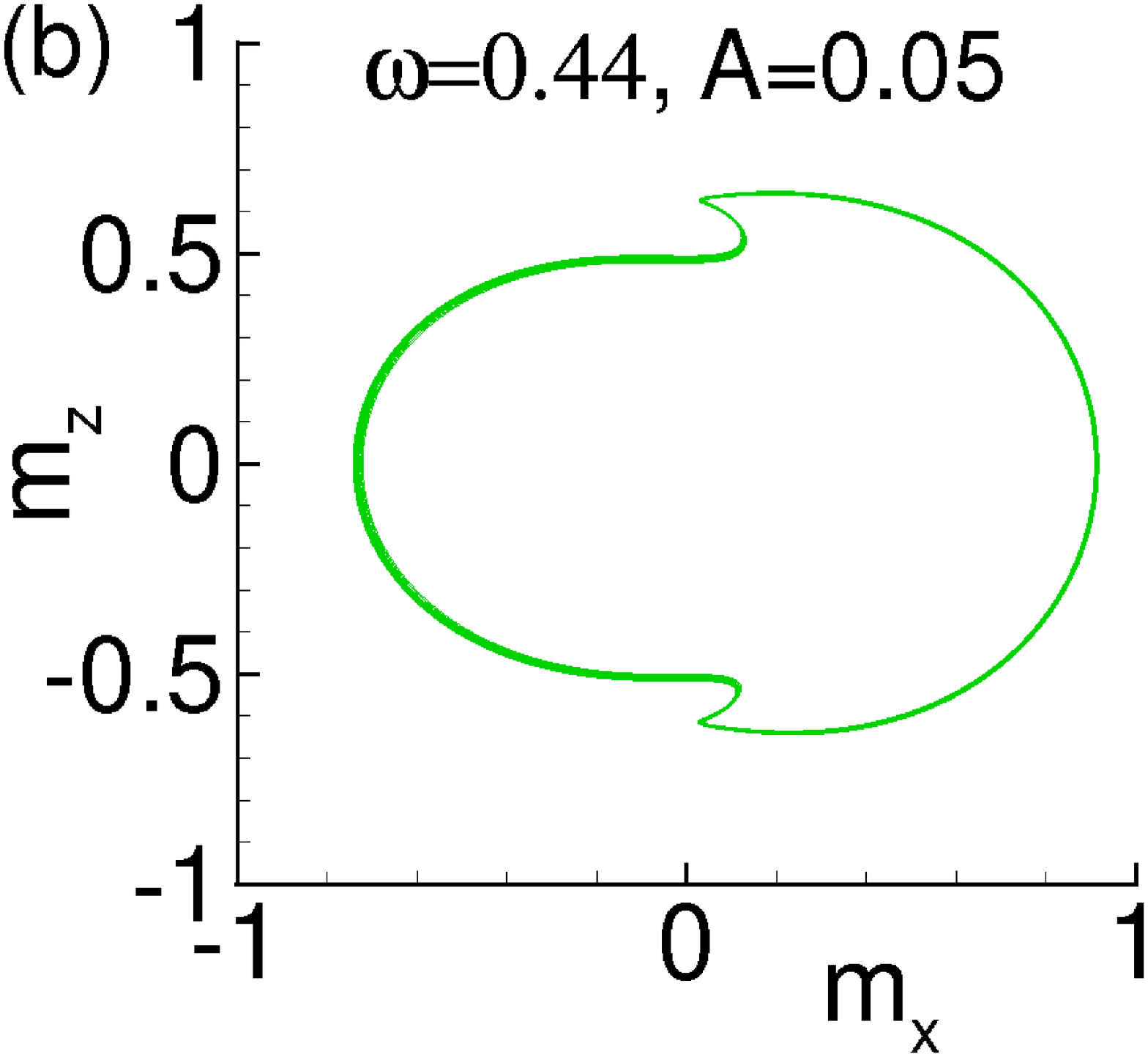}
\includegraphics[width=4.0cm]{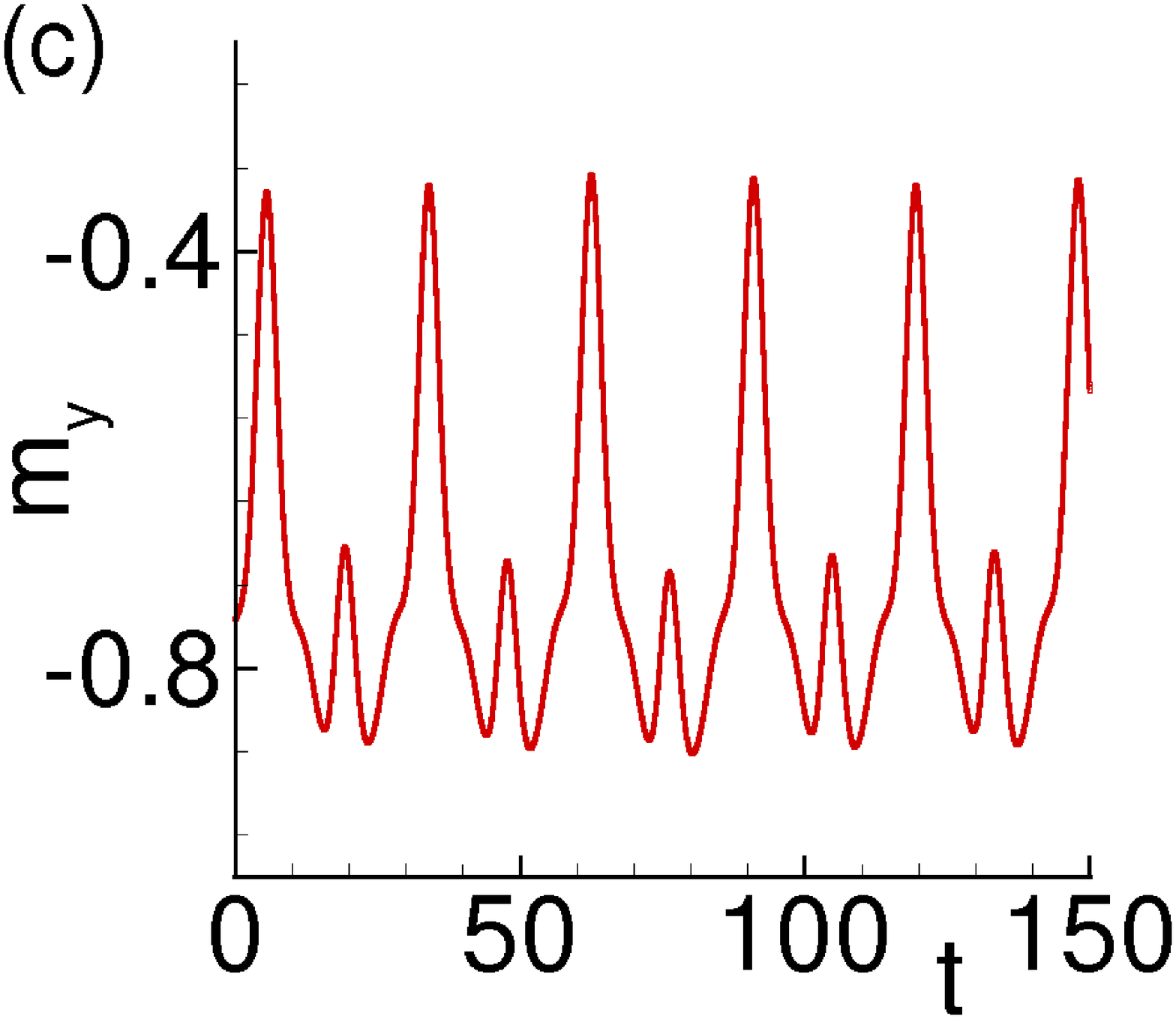}\includegraphics[width=4.0cm]{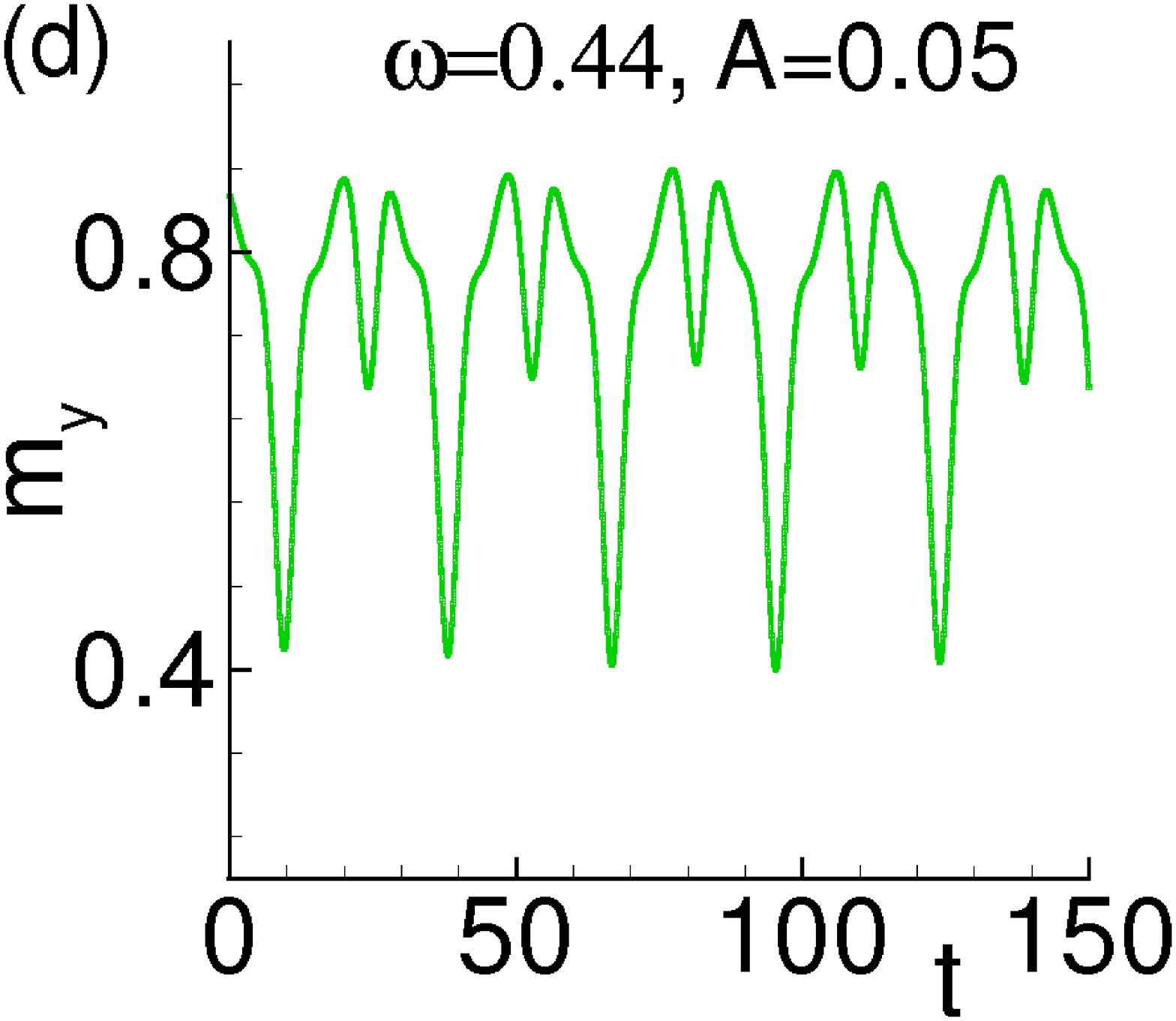}
\caption{Left-right transformation by changing the amplitude of
external radiation. (a) Dynamics of magnetic precession in $m_{z}$
-- $m_{x}$ plane without external radiation at $I=0.45$; (b) The
same under radiation with frequency $\omega=0.44$ and amplitude
$A=0.05$; (c) Time dependence of  $m_{y}$  without radiation at
$I=0.45$; (d) The same under radiation.} \label{7}
\end{figure}

Another central result of our work is to demonstrate that the
radiation may change the topology of magnetic precession. In
particular, we show the left-right transformation of
``mushroom''-type precession. As shown in Fig.\ \ref{7}(a,b), such a
change may be accomplished  by changing an amplitude of radiation at
a fixed DC drive current value $I=0.45$. This transformation is
related to a magnetization reversal from $-m_y$ to $+m_y$ as can be
seen from change in temporal dependence of $m_{y}(t)$ in the
presence of the external radiation as demonstrated in
Fig.~\ref{7}(c,d).

\section{ DC contribution to the Josephson current. }

As it was
stressed in Ref.~\cite{konschelle09,prb-ivc}, the Gilbert damping plays
an important role in the dynamics of S/F/S JJ.
It results in a DC contribution to the Josephson current
\begin{equation}
\label{current0}
I_{0}(\alpha)=\frac{\alpha G r^{2}\omega_{J}}{4}\bigg(\frac{1}{\Omega_{-}} + \frac{1}{\Omega_{+}}\bigg),
\end{equation}
with $\Omega_{\pm}=(\omega_{J}\pm1)^{2}+\alpha^{2}\omega_{J}^{2}$. As we see, this contribution depends on the spin-orbit interaction $r$ and relation of Josephson energy to magnetic energy $G$, and it is absent at $\alpha=0$.

\begin{figure}[h!]
\centering
\includegraphics[width=4.0cm]{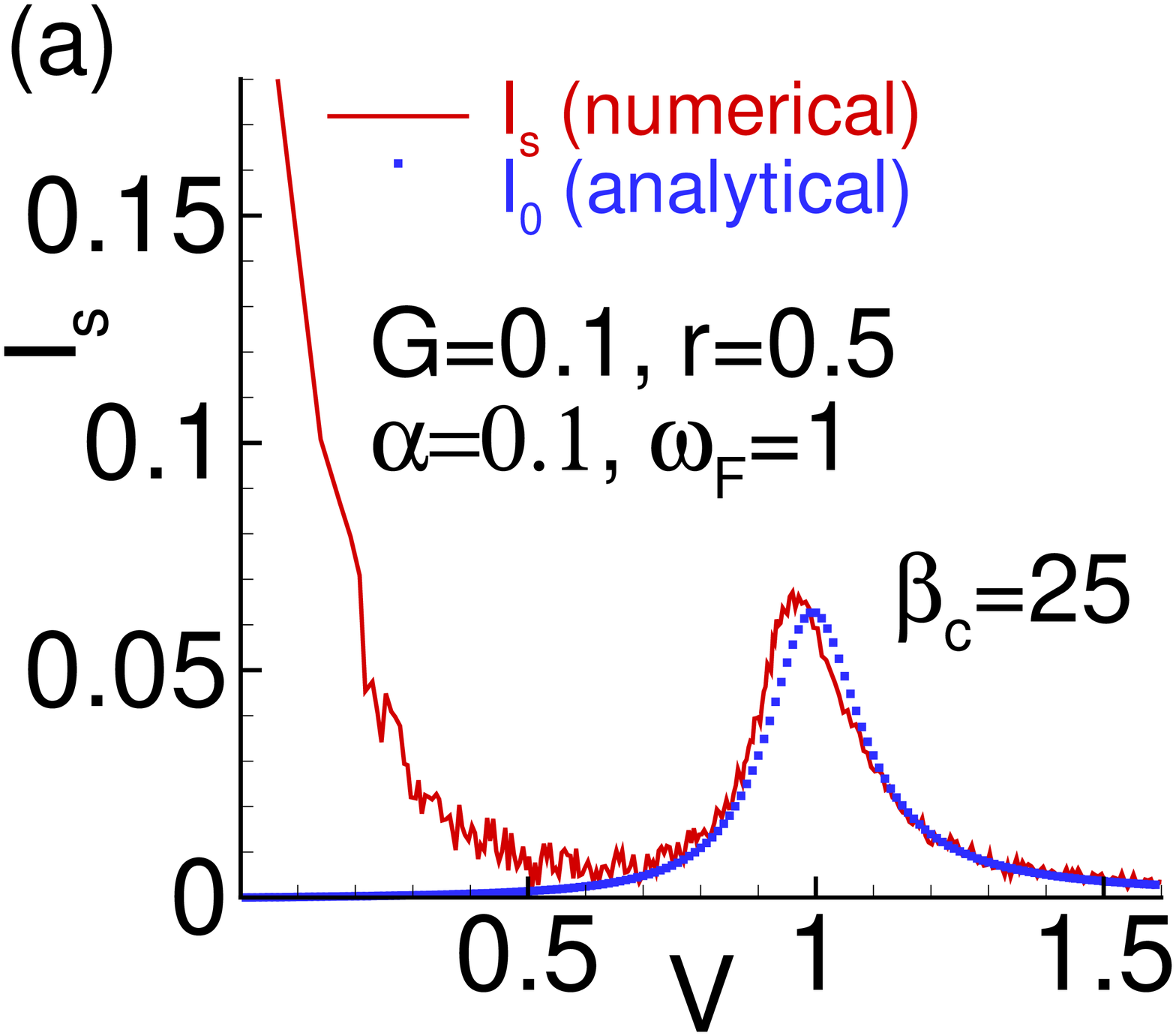}\includegraphics[width=4.0cm]{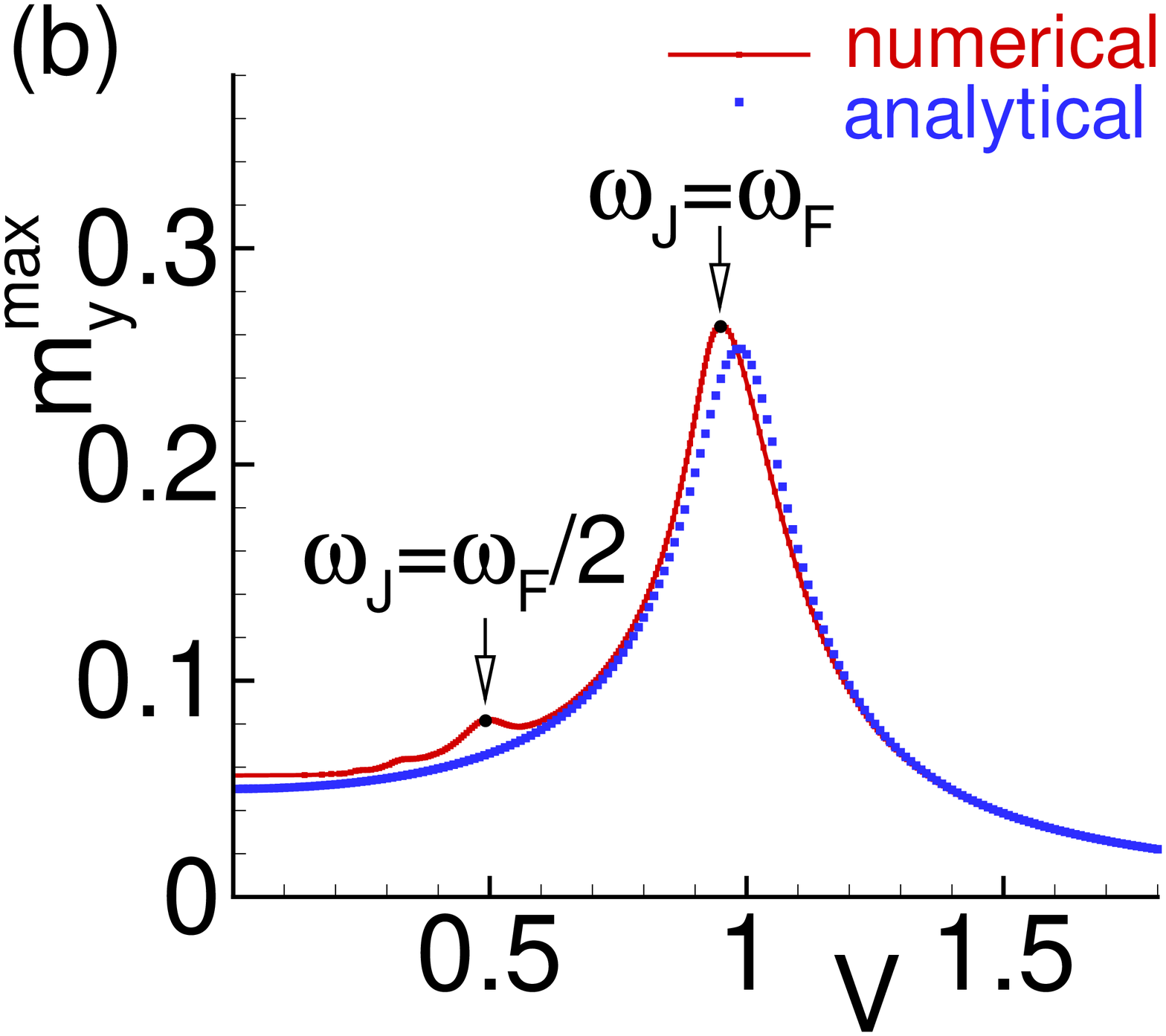}\\
\includegraphics[width=4.0cm]{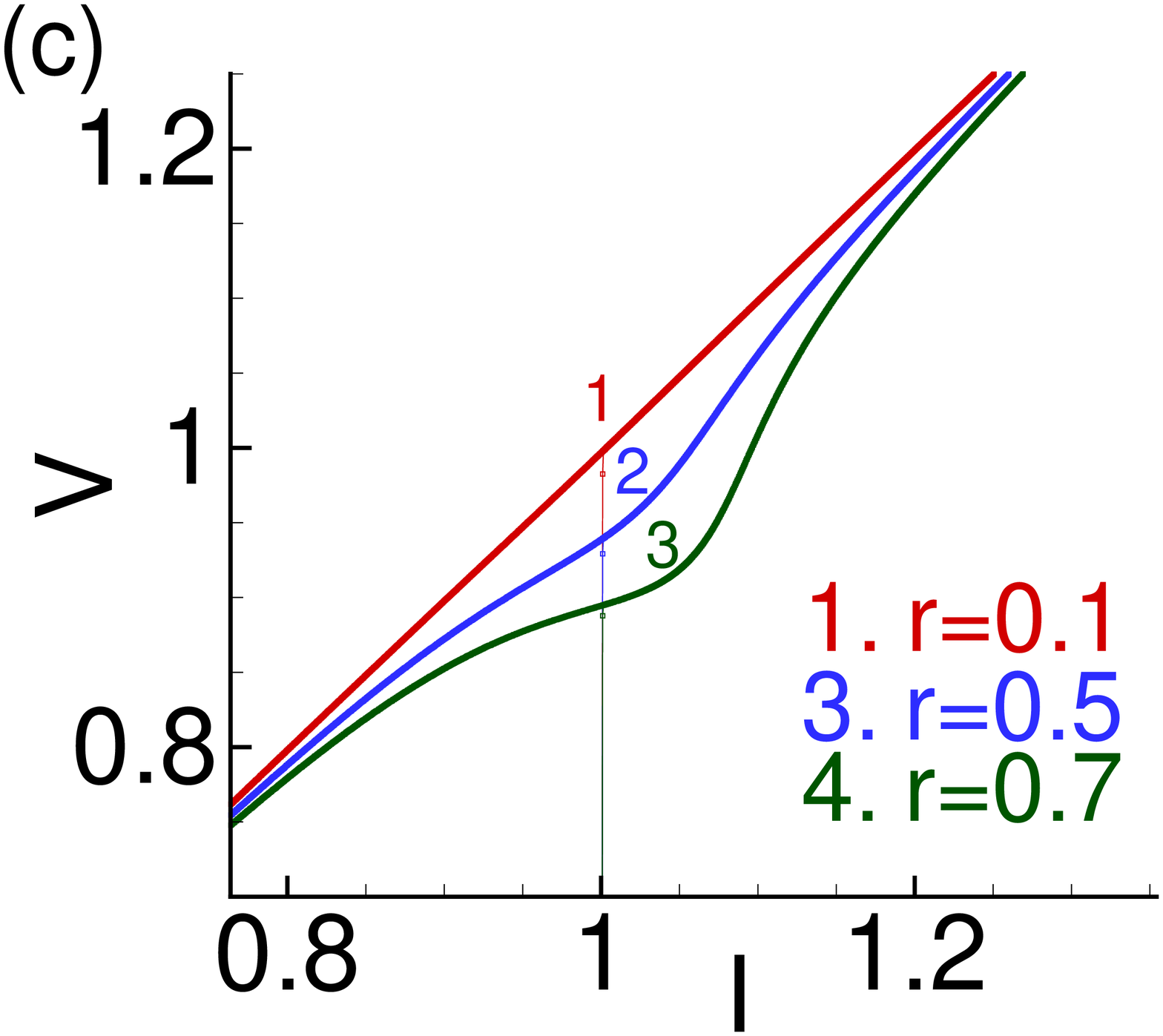}\includegraphics[width=4.0cm]{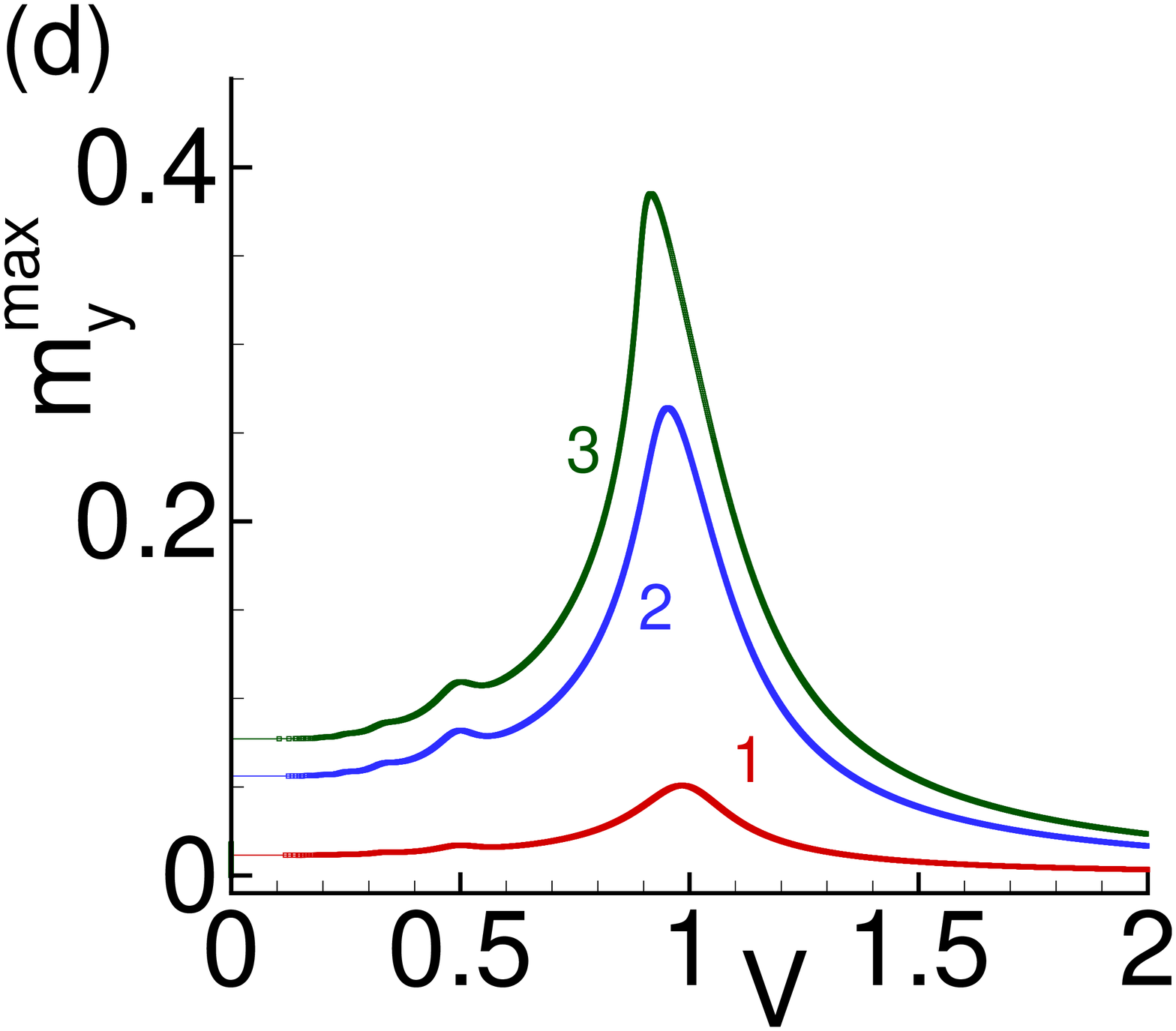}
\caption{ Demonstration of the ferromagnetic resonance
with sweeping bias current along IV-characteristics. (a) The voltage dependence of $I_{s}$ and analytical $I_{0}(\omega_{J})$; (b) The voltage dependence of $m^{max}_{y}$ and analytical $m_{y}(\omega_{J})$; (c)  The parts of IV--characteristics  of the $\varphi_0$ junction for $G=0.1$,$r=0.5$, $\alpha=0.1$, $\omega_{F}=1$ at different values of the spin-orbit interaction; (d) The voltage dependence of $m_{y}^{max}$ at different $r$.}
\label{8}
\end{figure}

The result of superconducting current simulation along IV-characteristics with sweeping bias current down is shown in Fig.\ \ref{8}(a). It presents the voltage dependence of $I_{s}(V)$ together with the analytical curve  for $I_{0}$,  according to (\ref{current0}). We see that in the resonance region the voltage dependence  of $I_{0}$ is in good agreement with  the result for superconducting current $I_{s}(V)$. The Gilbert damping leads to the damped ferromagnetic resonance at $\omega_J=\omega_F$ with corresponding analytical dependence~\cite{prb-ivc,konschelle09} for $m_y$
\begin{equation}
\label{solution2}
m_{y}(t)=\frac{\omega_{+}-\omega_{-}}{r}\sin\omega_{J} t-\frac{\alpha_{+}+\alpha_{-}}{r}\cos\omega_{J} t,
\end{equation}
where $\omega_{\pm}=\frac{Gr^{2}}{2}\frac{\omega_{J}\pm1}{\Omega_{\pm}}$ and $\alpha_{\pm}=\frac{Gr^{2}}{2}\frac{\alpha\omega_{J}} {\Omega_{\pm}}$
with $\Omega_{\pm}=(\omega_{J}\pm1)^{2}+\alpha^{2}\omega_{J}^{2}$. In Fig.\ref{8}(b), we plot this analytical dependence together with the maximal amplitude  $m_{y}^{max}$  calculated by the system of equations ($\ref{syseq}$) as a function of voltage. We see good
agrement of both results. We stress that numerical calculations do not use any approximations in comparison with analytical ones (where a weak coupling regime was used and considered the case $m_x,m_y<<1$ ), so the simulated dependence reflects additionally harmonic of the ferromagnetic resonance at $\omega_J=\omega_F/2$.

Based on the presented results, we may conclude that a variation of the Josephson junction and ferromagnetic layer parameters in the system with damping may lead to enough strong coupling between superconducting current and magnetization. Manifestation of such interaction in IV-characteristic of the $\varphi_0$ junction near the ferromagnetic resonance is presented in Fig.\ref{8}(c) where we show the parts of IV-characteristics of the $\varphi_0$ junction at three values of the spin-orbit interaction at $\omega_{F}=1$. The DC contribution to the Josephson current manifests itself as a deviation of IV-curve from the linear dependence in the resonance region.  The corresponding voltage dependencies of $m^{max}_y$ are shown in Fig.\ref{8}(d). The effect of the spin-orbit interaction on the resonance character of the presented  dependence might form a theoretical foundation for developing the experimental methods for determination of spin-orbit coupling intensity in the non-centrosymmetric  materials.

\section{ Conclusions. }
To summarize, we point out an intriguing opportunity to observe a
different type of magnetization trajectories by sweeping current  along the
IV-characteristics of the $\varphi_0$ junction due to a direct
coupling between the magnetic moment and the Josephson current. The prediction in Ref.\ \cite{konschelle09}
(which is verified in our numerical simulations) that the  DC
superconducting current in the presence of a constant voltage $V$
applied to the junction implies a dissipative regime can be easily
detected experimentally.  Good agreement between numerical
and analytical results found at the ferromagnetic resonance opening
wide opportunities for further manipulation of system parameters
and experimental verification of the magnetization
dynamics of the materials with strong spin-orbit coupling. This can be
easily achieved by applying external radiation to the setup used in
Ref.\ \cite{aprili}.

The appropriate candidate for the experimental verification of the obtained results might be a permalloy doped with $Pt$~\cite{hrabic16}. In this material  the parameter that characterizes the relative strength of the spin-orbit interaction
is $\upsilon_{so}/\upsilon_{F}\sim 1$. The $Pt$ at small doping ( up to $10\%$) did not influence significantly magnetic
properties of permalloy~\cite{hrabic16} and then we may expect $\upsilon_{so}/\upsilon_{F}$ to reach $0.1$
in this case also. If the length of the $F$ layer is of the order of the magnetic decaying length $\hbar/\upsilon_{F}=h$, i.e., $l\sim1$, we have $r\sim0.1$~\cite{apl17}. Another suitable candidate may be a $Pt/Co$ bilayer, ferromagnet without inversion symmetry like $MnSi$ or
$FeGe$. In this material the spin-orbit interaction can generate a $\varphi_{0}$
Josephson junction~\cite{buzdin08} with a finite ground phase difference.
The measurement of this phase difference may serve as an independent way
for the parameter $r$ evaluation~\cite{szombati}. The parameter $G$ has been
evaluated in Ref.\ \cite{konschelle09} for weak magnetic anisotropy of permalloy
$K\sim 4\times 10^{-5} K\AA^{-1}$ (see Ref. \cite{rusanov}) and S/F/S junction
with $l\sim1$ and $T_{c}\sim10K$ as $G\sim100$. For stronger anisotropy
we may expect $G\sim1$. The typical ferromagnetic resonance frequency is $\omega_{F}=10$ GHz which is accessible in the experiments. So, we may conclude, that our results can be tested experimentally.

A very rich physics is expected if the $\varphi_0$ Josephson junction is exposed to microwave radiation. Additionally to the features predicted in Ref.\cite{konschelle09}, particularly, an increase of the Shapiro steps amplitude due to spin-orbit coupling near the ferromagnetic resonance,  the appearance of the half-integer Shapiro steps, and precession of the magnetization vector with radiation frequency, we expect that an external electromagnetic field can control
qualitative features of the magnetic moment dynamics in a current
interval which corresponds to the Shapiro step. Moreover, as we demonstrated in this paper, such
radiation can also produce a topological transformation of
precession trajectories~\cite{cond-mat18}. We predict that the change in topology of the
magnetization dynamics would be observed in such systems as a
function of the amplitude of the applied electromagnetic radiation. We consider that the presented results might be used for developing novel experimental resonance methods for determination of the spin-orbit interaction in the non-centrosymmetric  materials.

\section{ Acknowledgements. }
The authors thank A. Mazanik,  I. Bobkova, A. Bobkov and A. Buzdin for helpful
discussion. The reported study was partially funded by the RFBR
research projects  18-02-00318, 18-52-45011-IND.  KS thanks DST for support through Indo-Russian grant INT/RUS/RFBR/314. Numerical calculations have been made in the framework of the RSF project 18-71-10095.

\end{document}